\documentclass[10pt]{article}
\usepackage[a4paper,total={6.00in, 9.0in}]{geometry}
\usepackage{moreverb,url}
\usepackage{amsmath,amssymb,graphicx}
\usepackage[colorlinks,bookmarksopen,bookmarksnumbered,citecolor=red,urlcolor=red]{hyperref}

\usepackage[super,comma,sort&compress]{natbib}
\makeatletter
\renewcommand\@biblabel[1]{#1.}
\makeatother

\newcommand\BibTeX{{\rmfamily B\kern-.05em \textsc{i\kern-.025em b}\kern-.08em
T\kern-.1667em\lower.7ex\hbox{E}\kern-.125emX}}

\begin{document}
\begin{center}
\textbf{\Large A quantile variant of the EM algorithm and its
       application to parameter estimation with interval data}

\medskip 
Chanseok Park

\medskip 
\textit{Department of Industrial Engineering, Pusan National University, Korea}
\end{center}

\begin{abstract}
The expectation-maximization (EM) algorithm is
a powerful computational technique
for finding the maximum likelihood estimates for parametric models
when the data are not  fully observed.
The EM is best suited for situations where the expectation
in each E-step and the maximization in each M-step are straightforward.
A difficulty with the implementation of the EM algorithm is that
each E-step requires the integration of the log-likelihood function
in closed form.
The explicit integration can be avoided by using
what is known as the Monte Carlo EM (MCEM) algorithm.  The MCEM uses a
random sample to estimate the integral at each E-step.  
However, the problem with the MCEM is that it often converges to the integral quite slowly
and the convergence behavior can also be unstable,
which causes a computational burden.
In this paper, we propose what we refer to as
the quantile variant of the EM (QEM) algorithm.
We prove that the proposed QEM method has an accuracy of $O(1/K^2)$ while
the MCEM method has an accuracy of  $O_p(1/\sqrt{K})$. 
Thus, the proposed QEM method possesses faster 
and more stable convergence properties when
compared with the MCEM algorithm. The improved performance is illustrated
through the numerical studies.
Several practical examples
illustrating its use in interval-censored data problems are also
provided.

\end{abstract}


\section{Introduction}
The analysis  
 of lifetime or failure time data has been of considerable
interest in many branches of applied engineering statistics 
including reliability engineering, biological sciences, etc. 
In reliability analysis, 
due to inherent limitations, or time and cost considerations on experiments.
The data are said to be censored when, for certain observations, 
only a lower or upper bound on the lifetime is available. 
Thus, there is partial information in the data set that still 
can be used in estimation  for reliability analysis.  
To obtain the parameter estimate, 
numerical optimization is often required to find the MLE.
However, ordinary numerical methods such as the Gauss-Seidel iterative method
and the Newton-Raphson gradient method may be very ineffective for complicated likelihood
functions and these methods 
can be sensitive to the choice of starting values used.
In this paper, unless otherwise specified, 
 ``MLE'' refers to the estimate obtained
by direct maximization of the likelihood function.

For censored sample problems, several approximations of the MLE and 
the best linear unbiased estimate (BLUE) have been studied 
instead of direct calculation of the MLE. 
For example, the problem of parameter estimation from censored samples
has been treated by several authors.
Gupta\cite{Gupta:1952} has studied the MLE and provided the BLUE
for Type-I and Type-II censored samples from a normal distribution.
Govindarajulu\cite{Govindarajulu:1966} has derived
the BLUE for a symmetrically
Type-II censored sample from a Laplace distribution only for sample size 
up to $n=20$.  
Balakrishnan\cite{Balakrishnan:1989} has given an approximation of the MLE
of the scale parameter of the Rayleigh distribution with censoring.
Hassanein et al.\cite{Hassanein/Saleh/Brown:1995} also has given a BLUE for 
a Type-II censored sample from Rayleigh distribution.
This BLUE, however, is limited to the case where the sample sizes are 
$n=5(1)25(5)45$ and the numbers of censored observations are $r=0,1,\ldots,n-2$, 
see Appendix F of Elsayed.\cite{Elsayed:2012}
Sultan\cite{Sultan:1997} has given an approximation of the MLE
for a Type-II censored sample from a normal distribution.
Balakrishnan\cite{Balakrishnan:1996} has given the
BLUE for a Type-II censored sample from a Laplace distribution.
The BLUE needs the coefficients  $a_i$ and $b_i$, which
were tabulated in Balakrishnan,\cite{Balakrishnan:1996} but
the table is provided only for sample size up to $n=20$.
In addition, the approximate MLE and the BLUE is not guaranteed 
to converge to the preferred MLE.
The methods above are also restricted to Type-I or Type-II (symmetric) censoring 
for sample size up to $n=20$ only.

The previously mentioned deficiencies can be overcome through the use of the EM algorithm.
However, in many practical problems, 
the implementation of the ordinary EM algorithm is very difficult because 
the expectation of the log-likelihood in the E-step can 
be quite complex or unavailable in closed form.
In order to avoid the explicit construction of the expectation in the E-step,
Wei and Tanner\cite{Wei/Tanner:1990a,Wei/Tanner:1990b} 
proposed the use of the Monte Carlo EM (MCEM) algorithm when the E-step is intractable. 
The MCEM algorithm uses Monte Carlo random sampling from
the conditional distribution in order to construct an empirical estimate of
the expected log-likelihood.  However, the MCEM algorithm often presents
difficulties because the convergence to the expected likelihood can
often be slow and unstable.
Therefore, we propose a quantile variant of the EM (QEM) algorithm
that constructs the empirical estimate of the expected log-likelihood
by non-random quantiles.  
The proposed variant is shown to have much faster convergence behavior 
and greater stability than
the MCEM while at the same time requiring smaller sample sizes.

Moreover, in many experiments, more general incomplete observations
are often encountered along with the fully observed data, where
incompleteness arises due to right-censoring, left-censoring, grouping,
quantal responses, etc.  
A general type of incomplete observations
is of interval form.  That is, a lifetime of a subject $X_i$ is specified
as $a_i \le X_i \le b_i$. 
We deal with computing the MLE
for this general form of incomplete data using the EM algorithm and its
variants, the MCEM and QEM algorithms.  
This interval form can handle right-censoring,
left-censoring, quantal responses and fully-observed observations.
This proposed method can also handle the data from intermittent inspection
which are referred to as \emph{grouped data}.
In the grouped data case, only the number of failures in each inspection period are
provided.
For example, the articles\cite{Seo/Yum:1993,Shapiro/Gulati:1998} 
provide an example using grouped data but they approximate the MLE 
and only consider 
the case where the lifetimes are exponentially distributed.
Nelson\cite{Nelson:1982} considers the maximum likelihood for grouped data 
but uses 
ordinary numerical methods which, as mentioned earlier, can often be problematic.
The attractiveness of our proposed method is
that it allows one to obtain the MLE using the QEM sequences under
a variety of distributional assumptions. We will illustrate that it is
easily applied to the cases described above and also provides more accurate
estimates.

\section{The EM and MCEM algorithms}
In this section, we give a brief introduction of the EM and MCEM algorithms.
Introduced by Dempster et al.,\cite{Dempster/Laird/Rubin:1977} the EM algorithm
is a powerful computational technique for finding  the MLE of parametric models
when there is no closed-form MLE, or the data are incomplete.
For more details about this EM algorithm, 
good references are Little and Rubin,\cite{Little/Rubin:2002} Tanner,\cite{Tanner:1996}
Schafer,\cite{Schafer:1997} and Hunter and Lange.\cite{Hunter/Lange:2004}

When the closed-form MLE from the likelihood function is not available,
numerical methods are required to find the maximizer (\emph{i.e.}, MLE).
However, ordinary numerical methods such as the Gauss-Seidel iterative
method and the Newton-Raphson gradient method may be very ineffective
for complicated likelihood functions and these methods can be sensitive
to the choice of starting values used.  In particular, if the likelihood
function is flat near its maximum, the methods will stop before reaching
the maximum.
These potential problems can be overcome by using the EM algorithm.

The EM algorithm consists of two iterative steps:
(i) the expectation step (E-step) and (ii) the maximization step (M-step).
The advantage of the EM algorithm is that it solves a difficult incomplete-data
problem 
by constructing two relatively straightforward steps.
The E-step of each iteration computes the conditional expectation of the
log-likelihood with respect to the incomplete data given the observed
data. The M-step of each iteration then obtains the maximizer of the
expected log-likelihood constructed in the E-step. 
Thus, the EM sequences repeatedly maximize the log-likelihood function
of the complete data given the incomplete data instead of maximizing 
the potentially complicated likelihood function of the incomplete data directly.
An additional advantage of this method compared to 
other direct optimization techniques is that it is very simple 
and it converges reliably.
In general, if it converges, it converges to a local maximum.
 Hence, in the case of the unimodal and concave likelihood function,
 the EM sequences converge to the global maximizer from any starting value.
We can employ this methodology for parameter estimation 
for interval-censored data 
because interval-censored data models are special cases 
of incomplete (missing) data models.

Here, we give a brief introduction of the EM and MCEM algorithms. 
Denote the vector of unknown parameters 
by ${\boldsymbol{\theta}}=(\theta_1,\ldots,\theta_p)$.
Then the complete-data likelihood is
$$
L^c({\boldsymbol{\theta}} | {\mathbf{x}})
= \prod_{i=1}^{n} f(x_i),
$$
where $\mathbf{x}=(x_1,\ldots,x_n)$ and we denote the observed part of $\mathbf{x}$ by
${\mathbf{y}}=(y_1,\ldots,y_m)$ and the incomplete (missing) part by
${\mathbf{z}}=(z_{m+1},\ldots,z_{n})$.
Denote the estimate at the $s$-th EM sequences by ${\boldsymbol{\theta}}^{(s)}$.
The EM algorithm consists of two distinct steps:
\begin{itemize}  
\item \textsf{E-step:}
      \texttt{Compute} $Q({\boldsymbol{\theta}}|{\boldsymbol{\theta}}^{(s)})$\\
where 
$Q({\boldsymbol{\theta}}|{\boldsymbol{\theta}}^{(s)})
   = \int\log L^c ({\boldsymbol{\theta}}|{\mathbf{y,z}}) \, 
  p({\mathbf{z}}|{\mathbf{y}},{\boldsymbol{\theta}}^{(s)})d{\mathbf{z}}$.

\bigskip
\item \textsf{M-step:}
      \texttt{Find}  ${\boldsymbol{\theta}}^{(s+1)}$ \\
which maximizes $Q({\boldsymbol{\theta}}|{\boldsymbol{\theta}}^{(s)})$ 
with respect to ${\boldsymbol{\theta}}$.
\end{itemize}  
\medskip

As stated earlier, the implementation of the E-step
in the EM algorithm can sometimes be quite difficult.  In order to avoid
this difficulty, 
Wei and Tanner\cite{Wei/Tanner:1990a,Wei/Tanner:1990b} 
proposed the MCEM algorithm.
In the MCEM, the expected log-likelihood in the E-step is approximated by using
Monte Carlo integration. By simulating $z_{m+1},\ldots,z_{n}$ from the conditional
distribution $p({\mathbf{z}}|{\mathbf{y}},{\boldsymbol{\theta}}^{(s)})$, 
the MCEM approximates the expected log-likelihood in the E-step. 
Let $K$ denote the number of samples used in the
 Monte Carlo integration of the MCEM and denote each simulated sample by
${\mathbf{z}}^{(k)}=(z_{m+1,k},\ldots,z_{n,k})$.
Then the Monte Carlo approximation of the expected log-likelihood is
\begin{equation}  
\widehat{Q}({\boldsymbol{\theta}}|{\boldsymbol{\theta}}^{(s)})
=\frac{1}{K}\sum_{k=1}^{K} 
 \log L^c ({\boldsymbol{\theta}}|{\mathbf{y}},{\mathbf{z}}^{(k)}).
\label{EQ:MCEM}
\end{equation}
This method where the E-step is changed to create an
empirical estimate of the expected log-likelihood is called the 
MCEM algorithm. Unfortunately, the major drawback to the
MCEM algorithm is that it can often be very slow because it requires a
large sample size for the empirical estimate to converge to the expected
likelihood. 
In addition, the values of the parameter estimation 
during each run of the MCEM algorithm 
can vary because random samples are used in the Monte Carlo integration. 
In fact, the dependence of the MCEM algorithm on random sampling implies that, 
even when using a large number of iterations, 
two identical runs of the MCEM algorithm can result in
different parameter estimates. These issues that arise due to the dependence
of the MCEM algorithm on random sampling are avoided in the QEM algorithm through
the use of deterministic sequences. 
In fact, random sampling is completely avoided in the QEM.

\section{The quantile variant of the EM algorithm}
The key idea underlying the QEM algorithm 
can be easily illustrated by the following example.
The data set in the example was first presented by Freireich et al.\cite{Freireich/etc:1963} 
and has since then been used very frequently for illustration in the reliability engineering 
and survival analysis literature.\cite{Leemis:1995,Leemis:2009,Cox/Oakes:1984}

\subsection{Illustrative example: length of remission of leukemia patients}
An experiment is conducted to determine the effect of a drug
named 6-mercaptopurine (6-MP) on leukemia remission times.
Twenty-one leukemia patients ($n=21$) are treated with 6-MP 
and the times of remission are recorded. 
There are nine individuals ($m=9$) for whom the remission time is fully observed,
and the remission times for the remaining twelve individuals are 
randomly censored on the right.
Letting a plus (+) denote a censored observation, the remission
times (in weeks) are: 
 6, 6,  6, $6^+$,  7,  $9^+$,  10,  $10^+$,  $11^+$,  13,  16, 
 $17^+$, $19^+$, $20^+$, 22, 23, $25^+$, $32^+$, $32^+$, $34^+$, $35^+$.

Assuming an exponential distribution for the lifetimes 
with the probability density function (pdf)
\[
f(x) = \frac{1}{\theta} e^{-x/\theta},
\]
we obtain the complete likelihood function 
\begin{align*}
\log L^c({\theta} | {\mathbf{y,z}})
&= -n\log\theta - \frac{1}{\theta}\sum_{i=1}^{m}y_i
 -\frac{1}{\theta}\sum_{i=m+1}^{n}z_i
\end{align*}
and the conditional pdf
\begin{align*}
p({\mathbf{z}}|{\mathbf{y}}, {\theta}^{(s)}, R_i)  
&=\prod_{i=m+1}^{n} p_{z_i}(z_i|\theta^{(s)}, R_i)  \\
&=\prod_{i=m+1}^{n} 
  \frac{1}{\theta^{(s)}} e^{-(z_i-R_i)/\theta^{(s)}}, 
\end{align*}
where $z_i>R_i$ and $R_i$ is a right-censoring time of the $i$-th test unit.
Using the above conditional pdf, we have the expected log-likelihood
\begin{align*}
&Q({\theta}|{\theta}^{(s)})  \\
&= \int\log L^{c}(\theta|{\mathbf{y,z}}) \;
    p({\mathbf{z}}|{\mathbf{y}}, {\theta}^{(s)}, R_i)  
    d{\mathbf{z}}  \\
&= -n\log\theta
   -\frac{1}{\theta}\sum_{i=1}^{m} y_i
   -\frac{1}{\theta}\sum_{i=m+1}^{n}\int z_i\;p_{z_i}(z_i|\theta^{(s)}, R_i) dz_i\\
&= -n\log\theta
   -(n-m) \frac{\theta^{(s)}}{\theta} 
   -\frac{1}{\theta}\sum_{i=1}^{m} y_i
   -\frac{1}{\theta}\sum_{i=m+1}^{n}R_i.
\end{align*}

In the Monte Carlo approximation, the term  
\[
E[z_i | \theta^{(s)}] = \int z_i p_{z_i}(z_i|\theta^{(s)}, R_i) dz_i
\]
is approximated by 
\begin{equation}
E[z_i | \theta^{(s)}]   
=\int z_i p_{z_i}(z_i|\theta^{(s)}, R_i) dz_i 
\approx 
\frac{1}{K}\sum_{k=1}^{K} z_{i,k},
\label{EQ:MCEMapprox}
\end{equation}
where a random sample $z_{i,k}$ is from 
\[
p_{z_i}(z_i|\theta^{(s)}, R_i)=\frac{1}{\theta^{(s)}} e^{-(z_i-R_i)/\theta^{(s)}}.  
\]
Then the Monte Carlo approximation of 
the expected log-likelihood is given by
\begin{align*}
\widehat{Q}({\theta}|{\theta}^{(s)})
= -n\log\theta
  -\frac{1}{\theta}\sum_{i=1}^{m} y_i
  -\frac{1}{\theta}\frac{1}{K}\sum_{k=1}^{K}\sum_{i=m+1}^{n} z_{i,k}.
\end{align*}

\begin{figure*}[t]
\centering
\includegraphics{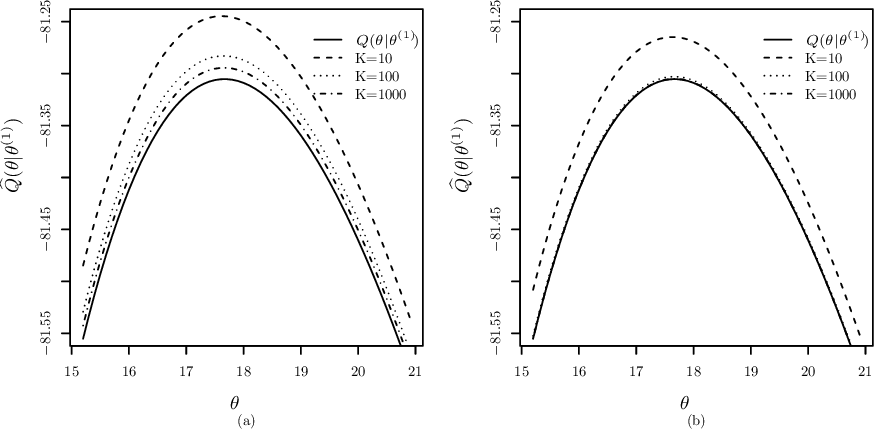}
\caption{The expected log-likelihood functions and approximations.
(a) Monte Carlo approximations. (b) Quantile approximations.\label{FIG:Qfunction}}  
\bigskip 
\end{figure*}

The key idea behind the QEM is that the approximation
above can be improved by using the quantile function.
Given the conditional pdf $p_{z_i}(z_{i,k}|\theta^{(s)},R_i)$, 
we denote the quantiles of $\xi_k$ as  
\begin{equation}
q_{i,k} = F^{-1}(\xi_k|\theta^{(s)}, R_i) = R_i - \theta^{(s)}\log(1-\xi_k).
\label{EQ:quantiles}
\end{equation}
One can choose $\xi_k$ from any form of the deterministic sequences such as 
$k/K$, $k/(K+1)$, $(k-\frac{1}{2})/K$, etc.
In this paper, we use $\xi_k=(k-\frac{1}{2})/K$ for $k=1,2,\ldots,K$.
By analogy with equation~(\ref{EQ:MCEMapprox}), we can approximate the term 
\[
E[z_i | \theta^{(s)}] = \int z_i p_{z_i}(z_i|\theta^{(s)}, R_i) dz_i. 
\]
Using the above quantiles $q_{i,k}$ in equation~(\ref{EQ:quantiles}) 
instead of a random sample $z_{i,k}$,
we have the following approximation 
\begin{equation}
E[z_i|\theta^{(s)}] 
= \int z_i p_{z_i}(z_i|\theta^{(s)}, R_i) dz_i 
\approx \frac{1}{K}\sum_{k=1}^K  q_{i,k}. 
\label{EQ:QEMapprox}
\end{equation}
It is noteworthy that a random sample $z_{i,k}$ in the Monte Carlo approximation 
can be generated by using the inverse transform algorithm.\cite{Ross:2013}
That is, the quantiles of a uniform random sample generate a random sample $z_{i,k}$.
However, the QEM uses the quantiles of the
deterministic sequences $\xi_k=(k-\frac{1}{2})/K$ 
which ensure faster and more stable convergence properties 
when compared with the MCEM.

Figure~\ref{FIG:Qfunction} presents 
the MCEM and QEM approximations of the expected log-likelihood functions 
for $K=10$~(dashed curve), 100~(dotted curve) and 1,000~(dot-dashed curve)
at the first step ($s=1$), 
along with the exact expected log-likelihood (solid curve).
The MCEM and QEM algorithms were run
with starting value $\theta^{(0)}=1$. 
As can be seen in Figure~\ref{FIG:Qfunction}, the MCEM and QEM both successfully converge
to the expected log-likelihood as $K$ gets larger. 
Note that the QEM is much closer to the true expected log-likelihood
for smaller values of $K$.
As afore-mentioned, it should be noted again that estimates based on the MCEM
can produce different values dependent on a random sample. 
Thus, the curves in Figure~\ref{FIG:Qfunction} (a) 
can change for each different random sample. On the other hand, 
the curves in Figure~\ref{FIG:Qfunction} (b) do not change because 
the QEM uses the deterministic sequences $\xi_k=(k-\frac{1}{2})/K$.

\begin{figure*}[t]
\centering
\includegraphics{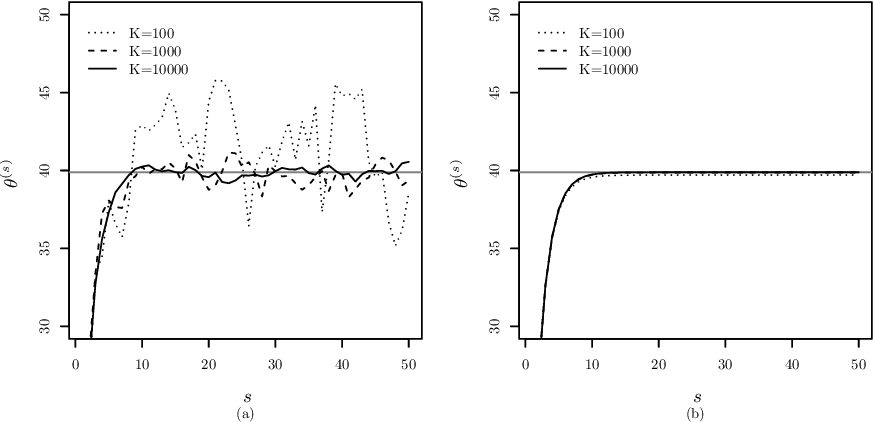}
\caption{\label{FIG:iteration}Successive parameter estimates
      using (a) the MCEM and  (b) the QEM.
       The horizontal solid lines indicate the MLE ($\hat{\theta}=39.89$).}
\end{figure*}

The plots of the parameter estimates at each value of $s$ for
the MCEM and QEM are shown in Figures~\ref{FIG:iteration} (a) and (b) 
respectively with the horizontal lines indicating the MLE ($\hat{\theta}=39.89$).
We used the starting value with $\theta^{(0)}=1$. 
The figures
clearly show that convergence behavior of the QEM is quite stable and
the number of steps required for convergence of the QEM is much smaller
than that of the MCEM. For example, using $K=100$ in the QEM results in
faster convergence than using $K=10,000$ in the MCEM.

\subsection{Convergence properties of the MCEM and QEM algorithms}
The two key questions are why the QEM is more stable and more accurate than the MCEM.
Both of the questions can be answered by considering 
the approximation in equation~(\ref{EQ:QEMapprox}) 
as an approximation to a Riemann-Stieltjes integral. 
For simplicity of presentation, we
only consider the case where ${\mathbf{z}}$ is one-dimensional 
but the same argument can be used in the case where ${\mathbf{z}}$ 
is multivariate. 
Denote $h({\theta},z)=\log L^{c}(\theta|{\mathbf{y}},z)$
and consider the following Riemann-Stieltjes sum 
\begin{equation}
\frac{1}{K} \sum_{k=1}^{K} h(\theta,F^{-1}\big( \xi_k \big)).
\label{EQ:Riemann}
\end{equation}
Note that in the limit as $K\to\infty$,  we have 
\begin{equation}
\frac{1}{K} \sum_{k=1}^{K} 
h(\theta,F^{-1}\big( \xi_k \big))  
\longrightarrow 
\int_0^1 h(\theta,F^{-1}(\xi)) d\xi.
\label{EQ:RiemannQEM}
\end{equation}
Using a change-of-variable integration technique with $z=F^{-1}(\xi)$, we have 
\[
  \int h(\theta,z) dF(z) = \int h(\theta,z) f(z) dz.
\]
Notice that the quantile approximation on the left-hand side 
of (\ref{EQ:RiemannQEM}) is a Riemann-Stieltjes sum which converges to the integral 
on the right-hand side of (\ref{EQ:RiemannQEM}). 
In our specific case, the integral represents 
the expected log-likelihood which therefore proves that the QEM converges.

The next step is to show why the QEM has better accuracy when compared with the MCEM.
With $\xi_k = (k-\frac{1}{2})/K$, the sum in equation~(\ref{EQ:Riemann}) is also known 
as the extended midpoint rule  which is well known to possess accuracy to the order of 
$O(1/K^2)$.\cite{Press/etc:2002} 
Specifically, it can be easily shown that
\begin{equation} \label{EQ:QEMaccuracy}
\int h(\theta,z) f(z) dz 
= \frac{1}{K} \sum_{k=1}^{K} h({\boldsymbol{\theta}},q_k) + O\Big(\frac{1}{K^2}\Big),
\end{equation} 
where $q_k = F^{-1}(\xi_k)$.
Thus, the accuracy of the integration in the E-step of the QEM is 
$O(1/K^{2})$.

On the other hand, the accuracy of the Monte Carlo approximation 
\[
\overline{h}_K = \frac{1}{K}\sum_{k=1}^{K} h(\theta, z_k)
\]
can be assessed as follows.
By the central limit theorem, we have
\begin{equation} \label{EQ:CLT}
\frac{ \sqrt{K}\Big\{\overline{h}_K-E( h(\theta,z))\Big\} }{\sqrt{{\mathrm{Var}}(h(\theta,z))}}
\stackrel{d}{\longrightarrow} N(0,1)
\end{equation}
which is accurate to the order of $O_p(1)$.
Using the weak law of large numbers, we have 
\[
\overline{h}_K \stackrel{p}{\longrightarrow} E(h(\theta,z)).
\]
Using this along with equation~(\ref{EQ:CLT}) results in 
\begin{equation} \label{EQ:MCEMaccuracy}
 \int h(\theta,z) f(z) dz 
=  \frac{1}{K}\sum_{k=1}^{K} h(\theta,z_k) + O_p\Big(\frac{1}{\sqrt{K}} \Big).
\end{equation}
Note that we have shown that the E-step of the QEM has
accuracy of deterministic $O(1/K^2)$ and the E-step of 
the MCEM has accuracy of probabilistic $O_p(1/\sqrt{K})$.
Therefore, the QEM has faster and more stable convergence properties 
compared to those of the MCEM.

We can generalize the above result as follows. 
In the E-step, using the quantiles instead of random samples,
we replace the Monte Carlo approximation of the expected log-likelihood 
in equation~(\ref{EQ:MCEM}) with the following quantile approximation
\[
\widehat{Q}(\theta|{\theta}^{(s)})
=\frac{1}{K}\sum_{k=1}^{K} 
 \log L^c (\theta|{\mathbf{y}},{\mathbf{q}}^{(k)}),
\]
where $\log L^c (\cdot)$ is the complete-data log-likelihood in the EM algorithm,
${\mathbf{q}}^{(k)}=(q_{m+1,k},\ldots,q_{n,k})$ 
with   $q_{i,k} = F^{-1}_{z_i}( \xi_k | {\theta}^{(s)})$,
and we used $\xi_k=(k-\frac{1}{2})/K$ as afore-mentioned. 

Note that the approximation of the expected log-likelihood 
in the proposed QEM method can be viewed as being similar
to a quasi-Monte Carlo approximation in the sense that the quasi-Monte
Carlo approximation also uses \textit{deterministic} sequences rather than a
\textit{random} sample.  
In fact, Niederreiter\cite{Niederreiter:1992} shows that there exist
such sequences in the normalized integration domain, which ensure accuracy 
on the order of $O(K^{-1} (\log K)^{d-1})$, where $d$ is the
dimension of the integration space.\cite{Robert/Casella:2005}
Thus, using the quasi-Monte Carlo sequences in the normalized integration domain,
one can improve the accuracy of the integration in the E-step 
of the MCEM algorithm which leads to accuracy to the order of $O(1/K^{1})$
with $d=1$. 
However, we should point out that the proposed QEM method
leads to accuracy to the order of $O(1/K^{2})$. 
Therefore, although using 
the quasi-Monte Carlo approximation can improve the convergence properties of the MCEM, 
the accuracy in that case will still be less than that of the proposed
QEM method.
Also, incorporating the quantiles from the proposed QEM method into 
the M-step to obtain the MLE is quite straightforward. 
Note also that, if the quasi-Monte Carlo sequences
in the normalized integration domain are used, this operation will not
have any relevance in the M-step in the sense that it still may be quite
difficult to obtain a closed-form solution for the maximization. 

Another way to approximate the expected log-likelihood is the use of a direct 
numerical integration in the E-step.
For example, instead of using the approximation 
\begin{align*} 
E[z_i | \theta^{(s)}] 
&= \int z_i \, p_{z_i}(z_i|\theta^{(s)}, R_i) dz_i  \\
&\approx (1/K)\sum_{k=1}^{K} z_{i,k}
\end{align*}
in equation~(\ref{EQ:MCEMapprox}), one may use 
\begin{align*} 
E[z_i | \theta^{(s)}] 
&= \int z_i \, p_{z_i}(z_i|\theta^{(s)}, R_i) dz_i  \\
&\approx \sum_{k=1}^{K} t_{i,k} \,  p_{z_i}(t_i|\theta^{(s)}, R_i) \Delta t_i
\end{align*}
where $\Delta t_i=t_{i,k} - t_{i,k-1}$, 
$t_{i,0}=a_i$ (the lower bound of the support of $z_i$), 
and $t_{i,K}=b_i$ (upper bound).
However, if the above direct numerical integration is used 
 instead of the MCEM approximation $(1/K)\sum_{k=1}^{K} z_{i,k}$ 
or the QEM approximation $(1/K)\sum_{k=1}^{K} q_{i,k}$, 
this can create a problem in the M-step because
this direct numerical integration includes the pdf term 
$p_{z_i}(t_i|\theta^{(s)}, R_i)$ in the sum.
Thus,  the integral becomes much more complex 
and this complexity can make it difficult or even impossible 
to find the closed-form maximizer in the M-step.
It should also be noted that the integrating domain of a direct numerical integration
is the same as the support of a random variable while the integrating domain
of the QEM method is always between zero and one as shown in equation~(\ref{EQ:RiemannQEM}).
If the support of a random variable is unbounded as is often the case 
in statistics, a numerical integration of an improper integral should be used; 
see Section 4.4 of Press et al.\cite{Press/etc:2002}
Improper integrals present serious challenges in numerical integration. 
In order to obtain reasonable accuracy using
numerical integration, great care needs to be taken and often advanced
methods need to be used.
Thus, the focus of the paper is to construct the
EM algorithm using the quantiles so that the closed-form maximizer in the 
M-step can be obtained in a straightforward manner.

\section{Likelihood construction}
In this section, we develop the likelihood functions which can be conveniently used 
for the EM, MCEM and QEM algorithms. 

The general form of an incomplete observation is often of interval form.
That is, the lifetime of a subject $X_i$ may not be observed  exactly, but is known
to fall in an interval, $a_i \le X_i \le b_i$. 
This interval form includes censored, grouped, quantal-response, and fully-observed observations.
For example, a lifetime is left-censored when $a_i=-\infty$ and
a lifetime is right-censored when $b_i=\infty$.
The lifetime is fully observed when $a_i = b_i$.

Suppose that ${\mathbf{x}}=(x_1,\ldots,x_n)$ are observations on 
random variables which are independent and identically distributed (iid)
and have a continuous distribution with pdf $f(x)$ and cumulative 
distribution function (cdf) $F(x)$.
Interval-censored data from experiments can be
conveniently represented by pairs $(w_i,\delta_i)$
with $w_i = [a_i,b_i]$,
\[
\delta_i = \left\{ \begin{array}{r@{\quad\mathrm{if}~}l}
                 0 &  a_i < b_i \\
                 1 &  a_i = b_i
                 \end{array} \right.
\qquad{\mathrm{for}}\quad i=1,\ldots,n,
\]
where $\delta_i$ is an indicator variable and $a_i$ and $b_i$ are lower and upper 
ends of interval observations of the $i$-th test unit, respectively. 
If $a_i = b_i$, then the lifetime of the $i$-th test unit is fully observed.
Denote the observed part of $\mathbf{x}=(x_1,\ldots,x_n)$ by
${\mathbf{y}}=(y_1,\ldots,y_m)$ and the incomplete (missing) part by
${\mathbf{z}}=(z_{m+1},\ldots,z_{n})$ with $a_i \le z_i \le b_i$.
Denote the vector of unknown parameters
by ${\boldsymbol{\theta}}=(\theta_1,\ldots,\theta_d)$.
Then ignoring a normalizing constant, we have the complete-data likelihood
function
\begin{equation} \label{EQ:likelihood0}
{L}^c({\boldsymbol{\theta}} | {\mathbf{y,z}})
\propto \prod_{i=1}^{n} f(x_i | {\boldsymbol{\theta}})
= \prod_{i=1}^{m} f(y_i | {\boldsymbol{\theta}}) \cdot 
 \prod_{i=m+1}^{n} f(z_i | {\boldsymbol{\theta}}),
\end{equation}
where the pdf of $z_i$ is given by
\begin{equation} \label{EQ:conditionalpdfzi}
p_{z_i}(z | {\boldsymbol{\theta}})
 = \frac{f(z| {\boldsymbol{\theta}})}{F(b_i |{\boldsymbol{\theta}})-F(a_i |{\boldsymbol{\theta}})
 },
\end{equation}
for $a_i < z < b_i$.

Integrating ${L}^c({\boldsymbol{\theta}} | {\mathbf{x}})$
with respect to ${\mathbf{z}}$, we obtain the observed-data likelihood
\begin{align*}
{L}({\boldsymbol{\theta}}|{\mathbf{y}})
&\propto \int {L}^c ({\boldsymbol{\theta}}|{\mathbf{y,z}}) d{\mathbf{z}} \\ 
&= \prod_{i=1}^{m} f(y_i | {\boldsymbol{\theta}})
  \prod_{i=m+1}^{n} \big\{F(b_i|{\boldsymbol{\theta}})-F(a_i|{\boldsymbol{\theta}})\big\},
\end{align*}
where an empty product is generally taken to be one.
Using the $(w_i,\delta_i)$ notation, we have
\begin{equation} \label{EQ:likelihood1}
{L}({\boldsymbol{\theta}} | {\mathbf{w}},{\boldsymbol{\delta}})
\propto \prod_{i=1}^{n} f(w_i |{\boldsymbol{\theta}})^{\delta_i} 
          \big\{F(b_i|{\boldsymbol{\theta}})-F(a_i|{\boldsymbol{\theta}})\big\}^{1-\delta_i},
\end{equation}
where ${\mathbf{w}}=(w_1,\ldots,w_n)$ and
      ${\boldsymbol{\delta}}=(\delta_1,\ldots,\delta_n)$.
Here, although we provided the likelihood function for the interval-data case,
it is easily extended to more general forms of incomplete data.
For more details, the reader is referred to Heitjan\cite{Heitjan:1989}
and Heitjan and Rubin.\cite{Heitjan/Rubin:1990}

Clearly, given the complexity of the likelihood,
the goal is to make an inference on ${\boldsymbol{\theta}}$ 
and the EM algorithm is a tool that
can be used to accomplish this goal.
Then the issue here is how to implement the EM algorithm
when there are interval-censored data in the sample.
By treating the interval-censored data as incomplete (missing) data, 
it is possible to write the complete-data
likelihood. 
This treatment allows one to fine the \emph{closed-form}\/ maximizer in the M-step.
For convenience,  assume that all the data are of interval form 
with $a_i \le w_i \le b_i$ and $a_i < b_i$.
Then the likelihood function in equation~(\ref{EQ:likelihood1}) can be rewritten as
\begin{equation} \label{EQ:likelihood2}
{L}({\boldsymbol{\theta}} | {\mathbf{w}})
\propto \prod_{i=1}^{n} \big\{ F(b_i | {\boldsymbol{\theta}})-F(a_i | {\boldsymbol{\theta}}) \big
\}.
\end{equation}
Then the complete-data likelihood function corresponding to 
equation~(\ref{EQ:likelihood2}) is given by 
\begin{displaymath} 
{L}^c({\boldsymbol{\theta}} | {\mathbf{y}}, {\mathbf{z}})
\propto   
 \prod_{i=1}^{n} f(z_i | {\boldsymbol{\theta}}),
\end{displaymath}
where the pdf of $z_i$ is given by
equation~(\ref{EQ:conditionalpdfzi}).
Using this result, we have the following $Q$-function in the E-step: 
\[
Q( {\boldsymbol{\theta}}|{\boldsymbol{\theta}}^{(s)} )
= \sum_{i=1}^{n} \int_{a_i}^{b_i}\log f(z_i|\boldsymbol{\theta})
      \cdot p_{z_i}(z_i|{\boldsymbol{\theta}^{(s)}})\, d
z_i.
\]

It is useful to consider the integral above when $b_i \to a_i$.
For notational convenience, 
omitting the subject index $i$ and letting $b=a+\epsilon$, we have
\begin{equation} \label{EQ:Qfunction1}
\int_{a}^{a+\epsilon}\log f(z|\boldsymbol{\theta}) \cdot p_{z}(z |{\boldsymbol{\theta}^{(s)}})\, dz. 
\end{equation}
It follows from integration by parts that the integral above becomes
\begin{equation} \label{EQ:Qfunction2}
\Big[ \log f(z|\boldsymbol{\theta}) P_z(z|{\boldsymbol{\theta}^{(s)}}) \Big]_a^{a+\epsilon}
 \!\!\! - \int_{a}^{a+\epsilon} \frac{f'(z|\boldsymbol{\theta})}{f(z|\boldsymbol{\theta})}
        P_z(z |{\boldsymbol{\theta}^{(s)}})\, dz, 
\end{equation}
where 
\begin{equation} \label{EQ:CDF-truncated}
P_z(z |{\boldsymbol{\theta}^{(s)}})
= \frac{F(z| \boldsymbol{\theta}^{(s)})}
  { F(a+\epsilon |\boldsymbol{\theta}^{(s)})-F(a|\boldsymbol{\theta}^{(s)}) }.
\end{equation}
Using equations~(\ref{EQ:Qfunction2}) and (\ref{EQ:CDF-truncated}), 
we can rewrite (\ref{EQ:Qfunction1}) as 
\begin{equation} \label{EQ:Qfunction3}
\frac{ A - B - C}%
{F(a+\epsilon |\boldsymbol{\theta}^{(s)})-F(a|\boldsymbol{\theta}^{(s)}) },
\end{equation}
where 
\begin{align*}
A &= \log f(a+\epsilon|\boldsymbol{\theta})F(a+\epsilon|\boldsymbol{\theta}^{(s)}),  \\
B &= \log f(a|\boldsymbol{\theta}) F(a|\boldsymbol{\theta}^{(s)}),
\intertext{and}
C &= \int_{a}^{a+\epsilon} \frac{f'(z|\boldsymbol{\theta})}{f(z|\boldsymbol{\theta})}
        \cdot F(z |{\boldsymbol{\theta}^{(s)}})\, dz.
\end{align*}
Applying l'Hospital rule to equation~(\ref{EQ:Qfunction3}), we obtain
\[
\lim_{\epsilon \to 0} 
\int_{a}^{a+\epsilon}\log f(z|\boldsymbol{\theta}) \cdot p_z(z |{\boldsymbol{\theta}^{(s)}})\, dz
= \log f(a | {\boldsymbol{\theta}}).
\]
Thus, in the case where all the lifetimes are fully observed, we simply
use the interval $[a_i, a_i]$ notation which
implies $[a_i, a_i+\epsilon]$ with the limit as $\epsilon\to0^{+}$.
Using this result, all the data points considered in
this paper can be viewed as data points in interval-data form without
requiring the use of the indicator variable $\delta_i$.

For notational convenience, we let $z_1=y_1, \ldots, z_m=y_m$.
Then the complete-data likelihood function corresponding to 
equation~(\ref{EQ:likelihood0}) becomes 
\begin{equation} 
{L}^c({\boldsymbol{\theta}} | {\mathbf{z}})
\propto  \prod_{i=1}^{n} f(z_i | {\boldsymbol{\theta}}),
\label{EQ:likelihood3}
\end{equation}
where ${\mathbf{z}}=(z_1, z_2, \ldots, z_n)$.
From now, unless otherwise specified,
${\mathbf{z}}$ refers to $(z_1, z_2, \ldots, z_n)$ instead of 
$(z_{m+1}, z_2, \ldots, z_n)$.
Thus, we use  equation~(\ref{EQ:likelihood3}) for the complete-data likelihood function 
rather than equation~(\ref{EQ:likelihood0}). 

For many distributions,
it is extremely difficult or even impossible to implement the EM algorithm 
with interval-censored data.
This is because, in the E-step, the $Q$-function does not integrate easily
and this causes computational difficulties in the M-step. In order to
avoid this problem, one can use the MCEM algorithm which 
reduces the difficulty in the E-step through the use of a Monte Carlo integration. 
As aforementioned, although it can make some problems tractable, 
the MCEM can be computationally very expensive
and often leads to unstable estimates.
Thus, we propose a quantile variant of the EM algorithm, the QEM, 
which alleviates the computational issues associated
with the MCEM algorithm and leads to more stable estimates.

Regardless of whether one uses EM, MCEM or QEM, a stopping criteria needs to 
be defined so that the algorithm converges after some number of iterations.
We define the stopping criteria as one in which the changes in successive estimates  
are relatively small compared to a defined precision $\epsilon$. 
For example, in the case
of the normal distribution, we can define the stopping criteria for
the QEM algorithm to occur when both 
\begin{align*}
\big|\mu^{(s+1)}-{\mu}^{(s)}\big| < \epsilon {\mu}^{(s+1)}  
\intertext{and}
\big|\sigma^{(s+1)}-{\sigma}^{(s)}\big| < \epsilon {\sigma}^{(s+1)},
\end{align*}
where $\epsilon$ is some small number which depends on one's desired
precision.
For other convergence criteria, the reader may refer to Press et al.\cite{Press/etc:2002}

In the section that follows, we maximize the likelihood function in 
equation~(\ref{EQ:likelihood1}) using the EM (when available), MCEM and QEM algorithms under
a variety of distributional assumptions.

\section{Parameter estimation}
In this section, we provide examples of parameter estimation
using the EM, MCEM and QEM algorithms under various distributional assumptions. 
Specifically, we consider the exponential, normal,
Laplace, Rayleigh and Weibull distributions in turn.

In the case where the exponential and normal distributions are assumed, 
the implementation of the EM algorithm is straightforward 
and there is actually no need to consider the MCEM or the QEM algorithms. 
Nevertheless, in order to compare the
performance of the MCEM and the QEM under those distributional assumptions, 
we include  the results of these approaches also.
Also, for the details involved in generating the EM sequences 
of the normal distribution with interval censoring,
the readers are referred to Lee and Park.\cite{Lee/Park:2006a}

Now, in the case where we assume that the
lifetimes have a Laplace distribution, the E-step computation in the EM
algorithm is extremely complex so the  MCEM and QEM are more appropriate
and we expect the QEM to outperform the MCEM.
Finally, when the  Rayleigh and Weibull distributions are assumed for
the lifetimes, the expected log-likelihood in the E-step of the EM does not have an
explicit integration so it is not possible to apply the EM algorithm in
these cases.

As aforementioned, it is noteworthy that the QEM sequences are easily obtained 
by replacing a random sample ${\mathbf{z}}^{(k)}$ in the MCEM sequences 
with quantile sequences ${\mathbf{q}}^{(k)}$.

\subsection{Exponential distribution}
We assume that the random variables $z_i$ are iid exponential random variables 
with the pdf given by $f(z|\lambda) = \lambda \exp(-\lambda z)$.
Using equation~(\ref{EQ:likelihood3}), 
we obtain the complete-data log-likelihood of ${\lambda}$
\[
\log L^c(\lambda | \mathbf{z} ) = \sum_{i=1}^{n} ( \log\lambda - \lambda z_i ),
\]
where the pdf of $z_i$ is given by
\begin{equation}  \label{EQ:ExpoTruncation}
p_{z_i}(z | \lambda )
 = \frac{\lambda \exp(-\lambda z)}{ \exp(-\lambda a_i) - \exp(-\lambda b_i) },
\end{equation}  
for $a_i < z < b_i$.
When $a_i=b_i$, the above random variables $z_i$ degenerate at $z_i=a_i$.

\begin{itemize}
\item \textsf{E-step:} \\
When $a_i < b_i$, the $Q(\cdot)$ function is given by
\begin{align*}
Q( \lambda | \lambda^{(s)} ) 
&= \int\log L^c(\lambda | {\mathbf{z}}) \, 
            p({\mathbf{z}} | \lambda^{(s)}) d{\mathbf{z}} \\
&= n\log\lambda - \lambda \sum_{i=1}^{n} A_{i}^{(s)},
\end{align*}
where   
\[
p({\mathbf{z}} | \lambda^{(s)}) 
 = \prod_{i=1}^{n} p_{z_i}(z_i | \lambda^{(s)} ) 
\]
and
\begin{align*}
A_{i}^{(s)} 
&= E[z_i |\lambda^{(s)}] 
 = \int_{a_i}^{b_i} \!\!\! z\cdot p_{z_i}(z | \lambda^{(s)} ) \, dz  \\
&= \frac{ a_i\exp(-\lambda^{(s)} a_i) - b_i\exp(-\lambda^{(s)} b_i)}
   { \exp(-\lambda^{(s)} a_i) -\exp(-\lambda^{(s)} b_i) } + \frac{1}{\lambda^{(s)}}.
\end{align*}
Note that  when $a_i=b_i$, we have $A_{i}^{(s)}=a_i$. 

\bigskip
\item \textsf{M-step:} \\
Differentiating
$Q(\lambda| \lambda^{(s)})$
with respect to $\lambda$ and setting this to zero, we obtain
\[
\frac{\partial Q(\lambda | \lambda^{(s)})}{\partial\lambda}
= \frac{n}{\lambda} -  \sum_{i=1}^{n} A_{i}^{(s)} = 0.
\]
Solving for $\lambda$, we obtain the $(s+1)$st EM sequence in the \mbox{M-step}
\begin{equation}  \label{EQ:EMexpo}
\lambda^{(s+1)} = \frac{n}{ \sum_{i=1}^{n}A_{i}^{(s)} }.
\end{equation}
\end{itemize}

If we instead use the MCEM algorithm by simulating 
$z_{1},\ldots,z_{n}$ from the truncated exponential distribution
$p({\mathbf{z}}|{\boldsymbol{\theta}}^{(s)})$,
we then obtain the MCEM sequences
\[
\lambda^{(s+1)} = \frac{n}{ \sum_{i=1}^{n} \frac{1}{K}\sum_{k=1}^{K} z_{i,k} },
\]
where $z_{i,k}$ for $k=1,2,\ldots,K$ are from the truncated exponential distribution 
$p_{z_i}(z|\lambda^{(s)})$ defined in equation~(\ref{EQ:ExpoTruncation}). 
On the other hand, if we use the QEM algorithm by quantiling, 
we then obtain the QEM sequences
\[
\lambda^{(s+1)} = \frac{n}{ \sum_{i=1}^{n} \frac{1}{K}\sum_{k=1}^{K} q_{i,k} },
\]
where $q_{i,k} = F_{z_i}^{-1}( \xi_k | {\lambda}^{(s)})$
and $\xi_k=(k-\frac{1}{2})/K$.  It is immediate from equation~(\ref{EQ:ExpoTruncation}) that 
\[
F_{z_i} (z|\lambda) 
 =\frac{\exp(-\lambda a_i)-\exp(-\lambda z)}{ \exp(-\lambda a_i)-\exp(-\lambda b_i)},
\]
for $a_i < z < b_i$, $F_{z_i} (z|\lambda)=0$ for $z\le a_i$,
and  $F_{z_i} (z|\lambda)=1$ for $z \ge b_i$.
Thus, the quantile sequences are explicitly obtained as 
\begin{align*}
q_{i,k}  &= \frac{1}{{\lambda}^{(s)}} \times   \\
 &\qquad \log\Big[ \frac{1}{ (1-\xi_k)\exp(-\lambda^{(s)} a_i) 
       + \xi_k \exp(-\lambda^{(s)} b_i) } \Big] .
\end{align*}

It is of interest to consider the case where the data are right-censored.
In this special case, the closed-form MLE is known.
If the data are fully observed 
(\emph{i.e.}, $w_i=[a_i,a_i]$) for $i=1,2,\ldots,r$,
it is easily seen from l'Hospital rule that $A_{i}^{(s)}=a_i$.
If the observation is right-censored 
(\emph{i.e.}, $w_i=[a_i,\infty]$) for $i=r+1,\ldots,n$,
we have $A_{i}^{(s)}=a_i + 1/\lambda^{(s)}$.
Substituting these results into equation~(\ref{EQ:EMexpo}) leads to
\begin{equation}  \label{EQ:EMexpo2}
\lambda^{(s+1)} = \frac{n}{ \sum_{i=1}^{n} a_i + (n-r)/\lambda^{(s)} }.
\end{equation}
Note that solving the stationary-point equation
$\hat{\lambda} = \lambda^{(s+1)}=\lambda^{(s)}$ of equation~(\ref{EQ:EMexpo2}) gives
\[
\hat{\lambda} = \frac{r}{ \sum_{i=1}^{n} a_i }.
\]
As expected, the results is identical to the well-known closed-form MLE 
in the right-censored data case.

\subsection{Normal distribution}
We assume that the random variables $z_i$ are iid normal random variables
with parameter vector $\boldsymbol{\theta} = (\mu, \sigma)$.
Using equation~(\ref{EQ:likelihood3}), 
we obtain the complete-data log-likelihood of~$\boldsymbol{\theta}$
\begin{align*}
\log L^c({\boldsymbol{\theta}} | \mathbf{z})
&\propto - \frac{n}{2}\log\sigma^2  - \frac{n}{2\sigma^2} \mu^2 \\
&\qquad - \frac{1}{2\sigma^2} \Big\{\sum_{i=1}^{n}\!z_i^2-2\mu\sum_{i=1}^{n}\!z_i \Big\},
\end{align*}
where the pdf of $z_i$ is given by
\begin{equation}  
p_{z_i}(z| {\boldsymbol{\theta}} )
 =  \frac{\frac{1}{\sigma} \phi(\frac{z-\mu}{\sigma})}
       {\Phi( \frac{b_i-\mu}{\sigma}) - \Phi( \frac{a_i-\mu}{\sigma})},
\label{EQ:NormalTruncation}
\end{equation}
for $a_i < z < b_i$.
Similarly as before, if $a_i=b_i$, then the random variables $z_i$ degenerate 
at $z_i = a_i$. 

\begin{itemize}
\item \textsf{E-step:} \\
Denote the estimate of $\boldsymbol{\theta}$ at the $s$-th EM sequence by
$\boldsymbol{\theta}^{(s)} = (\mu^{(s)}, \sigma^{(s)})$.
Ignoring constant terms, we have 
\begin{align*}
Q({\boldsymbol{\theta}}|{\boldsymbol{\theta}}^{(s)})
&= \int\log L^c({\boldsymbol{\theta}} | {\mathbf{z}}) \, 
            p({\mathbf{z}} | {\boldsymbol{\theta}}^{(s)}) d{\mathbf{z}} \\
&= -\frac{n}{2}\log\sigma^2 - \frac{n}{2\sigma^2}\mu^2
  -\frac{1}{2\sigma^2} \sum_{i=1}^{n} A_i^{(s)} \\
&\qquad + \frac{\mu}{\sigma^2} \sum_{i=1}^{n} B_i^{(s)},
\end{align*}
where 
\( p({\mathbf{z}} |   {\boldsymbol{\theta}}^{(s)}) 
 = \prod_{i=1}^{n} p_{z_i}(z_i | {\boldsymbol{\theta}}^{(s)}  ) \), 
 $A_i^{(s)}=E[z_i^2|{\boldsymbol{\theta}}^{(s)}]$ and
 $B_i^{(s)}=E[z_i|{\boldsymbol{\theta}}^{(s)}]$.
Using the following integral identities
\begin{align*}
\int\frac{z}{\sigma}\phi(\frac{z-\mu}{\sigma})dz
     &= \mu \Phi(\frac{z-\mu}{\sigma})-\sigma\phi(\frac{z-\mu}{\sigma})
\intertext{and}
\int\frac{z^2}{\sigma}\phi(\frac{z-\mu}{\sigma})dz
    &= (\mu^2+\sigma^2)\Phi(\frac{z-\mu}{\sigma}) \\
    &\qquad -\sigma(\mu+z)\phi(\frac{z-\mu}{\sigma}),
\end{align*}
we obtain 
\begin{align*}
A_i^{(s)} &=  \{\mu^{(s)}\}^2 + \{\sigma^{(s)}\}^2 - \sigma^{(s)} \times  \\
& \frac{(\mu^{(s)}+b_i)\phi(\frac{b_i-\mu^{(s)}}{\sigma^{(s)}})-(\mu^{(s)}+a_i)\phi(\frac{a_i-\mu^{(s)}}{\sigma^{(s)}}) }
                  {\Phi(\frac{b_i-\mu^{(s)}}{\sigma^{(s)}})-\Phi(\frac{a_i-\mu^{(s)}}{\sigma^{(s)}})}
\intertext{and}
B_i^{(s)} &= \mu^{(s)}
        - \sigma^{(s)} \times 
      \frac{\phi(\frac{b_i-\mu^{(s)}}{\sigma^{(s)}})-\phi(\frac{a_i-\mu^{(s)}}{\sigma^{(s)}})}
         {\Phi(\frac{b_i-\mu^{(s)}}{\sigma^{(s)}})-\Phi(\frac{a_i-\mu^{(s)}}{\sigma^{(s)}})},
\end{align*}
where  $a_i < b_i$.
It should be noted that for $a_i=b_i$ we have $A_i^{(s)}=a_i^2$ and $B_i^{(s)}=a_i$. 

\bigskip
\item \textsf{M-step:} \\
Differentiating the expected log-likelihood
$Q({\boldsymbol{\theta}}|{\boldsymbol{\theta}}^{(s)})$
with respect to $\mu$ and $\sigma^2$ and  solving for
$\mu$ and $\sigma^2$, we obtain the EM sequences
\begin{align}
\mu^{(s+1)}   &= \frac{1}{n} \sum_{i=1}^{n} B_i^{(s)}, \label{EQ:EMnormal1} 
\intertext{and}
{\sigma^2}^{(s+1)}  &= \frac{1}{n} \sum_{i=1}^{n}  A_i^{(s)}
            - \big\{ \mu^{(s+1)}  \big\}^2. \label{EQ:EMnormal2}
\end{align}
\end{itemize}

If we instead use the MCEM algorithm by simulating 
$z_{1},\ldots,z_{n}$ from the truncated normal distribution 
$p({\mathbf{z}}|{\boldsymbol{\theta}}^{(s)})$,
we then obtain the MCEM sequences 
\begin{align} 
\mu^{(s+1)}  & = \frac{1}{n} \sum_{i=1}^{n} \frac{1}{K} \sum_{k=1}^{K} z_{i,k}, 
                 \label{EQ:MCEMnormal1} 
\intertext{and}
{\sigma^2}^{(s+1)}&= \frac{1}{n} \sum_{i=1}^{n} \frac{1}{K} \sum_{k=1}^{K} z_{i,k}^2 
                    - \big\{ \mu^{(s+1)}  \big\}^2, 
              \label{EQ:MCEMnormal2} 
\end{align}
where $z_{i,k}$ are from the truncated normal distribution 
$p_{z_i}({z_{i,k}}|\mu^{(s)},\sigma^{(s)})$ defined in equation~(\ref{EQ:NormalTruncation}). 
Note that the QEM algorithm is easily obtained by quantiling 
$z_{1},\ldots,z_{n}$.
As illustrated in the exponential case, the quantiles are easily obtained using
$q_{i,k} = F^{-1}_{z_i}( \xi_k | \mu^{(s)},\sigma^{(s)} )$.
Thus, replacing $z_{i,k}$ in equations~(\ref{EQ:MCEMnormal1}) and (\ref{EQ:MCEMnormal2}) with
$q_{i,k}$, we can obtain the QEM sequences. 

\subsection{Laplace distribution}
We assume that the random variables $z_i$ are iid Laplace random variables with 
parameter $\boldsymbol{\theta}=(\mu,\sigma)$ whose pdf is given by 
\[
f(x| \mu,\sigma)
  = \frac{1}{2\sigma} \exp\Big(-\frac{|x-\mu|}{\sigma}\Big).
\]
Using equation~(\ref{EQ:likelihood3}), 
we have the complete-data log-likelihood of 
$\boldsymbol{\theta}$
\begin{align*}
\log {L}^c({\boldsymbol{\theta}} | {\mathbf{z}}) 
&= C - n\log\sigma
- \frac{1}{\sigma} \sum_{i=1}^{m} |y_i-\mu| \\
&\quad - \frac{1}{\sigma} \sum_{i=m+1}^{n} |z_i-\mu|,
\end{align*}
where the pdf of $z_i$ is given by
\begin{equation} \label{EQ:LaplaceTruncation}
p_{z_i}(z | \boldsymbol{\theta} )
 = \frac{ f(z|\boldsymbol{\theta}) }{ F(b_i|\boldsymbol{\theta}) - F(a_i|\boldsymbol{\theta}) }
\end{equation}
for $a_i < z < b_i$.
Similarly as before, if $a_i=b_i$, then the random variables $z_i$ 
degenerate at $z_i=a_i$.

\begin{itemize}
\item \textsf{E-step:} \\
At the $s$-th step in the EM sequence denoted by 
${\boldsymbol{\theta}}^{(s)}=(\mu^{(s)},\sigma^{(s)})$, 
we have the expected log-likelihood
\begin{align*}
&Q({\boldsymbol{\theta}}|{\boldsymbol{\theta}}^{(s)})  \\
&= \int\log L^c({\boldsymbol{\theta}} | {\mathbf{z}}) \, 
            p({\mathbf{z}} | {\boldsymbol{\theta}}^{(s)}) d{\mathbf{z}} \\
&= C - n\log\sigma
- \frac{1}{\sigma} \sum_{i=1}^{n}\int_{a_i}^{b_i} |z_i-\mu| \,
            f({z_i}|{\boldsymbol{\theta}}^{(s)}) d{z_i}.
\end{align*}
Note that integrating the third term in the
expression above is extremely complex. We can avoid this difficulty by
using the MCEM algorithm or the QEM algorithm. Using the standard MCEM
technique given $K$ samples, the approximate expected log-likelihood
becomes
\begin{align}
&\widehat{Q}({\boldsymbol{\theta}}|{\boldsymbol{\theta}}^{(s)})  \notag \\
&=\frac{1}{K} \sum_{k=1}^{K} \log L^c({\boldsymbol{\theta}} | {\mathbf{z}}^{(k)}) 
       \notag \\
&=C - n\log\sigma - \frac{1}{\sigma} \sum_{i=1}^{n}   \frac{1}{K}\sum_{k=1}^{K}
    \big| z_{i,k}-\mu \big|, \label{EQ:LaplaceQhat}
\end{align}
where $\mathbf{z}^{(k)}=(z_{1,k}, z_{2,k}, \ldots, z_{n,k})$.
Therefore, we can estimate the expected log-likelihood by generating 
$z_{i,k}$ for $k=1,2,\ldots,K$ from $p_{z_i}(z|{\boldsymbol{\theta}}^{(s)})$ 
defined in equation~(\ref{EQ:LaplaceTruncation}).
Then by replacing $z_{i,k}$ in equation~(\ref{EQ:LaplaceQhat}) with the quantiles 
$q_{i,k} = F^{-1}_{z_i}( \xi_k | \mu^{(s)},\sigma^{(s)} )$, 
the E-step for the QEM algorithm is easily obtained.

\bigskip
\item \textsf{M-step:} \\
It is straightforward to obtain the MCEM and QEM sequences 
which maximize equation~(\ref{EQ:LaplaceQhat})
\begin{align}  
\mu^{(s+1)}
&= {\mathrm{median}}
  ( {\mathbf{z}}^{(1)},\ldots,{\mathbf{z}}^{(K)}) \label{EQ:MCEMLaplace1} 
\intertext{and}
\sigma^{(s+1)}
&= \frac{1}{n} \sum_{i=1}^{n} \frac{1}{K}\sum_{k=1}^{K} \big|z_{i,k}-\mu^{(s+1)}\big|. 
                                     \label{EQ:MCEMLaplace2}
\end{align}
Again, replacing $z_{i,k}$ in equations~(\ref{EQ:MCEMLaplace1}) and (\ref{EQ:MCEMLaplace2}) with
the quantiles $q_{i,k}$ provides the QEM sequences. 
\end{itemize}

Note that  if the direct numerical integration is used instead of 
the MCEM or QEM approximation, the approximate expected log-likelihood becomes
\begin{align}  
&\widehat{Q}({\boldsymbol{\theta}}|{\boldsymbol{\theta}}^{(s)}) \notag \\
&= C - n\log\sigma - \frac{1}{\sigma} \sum_{i=1}^{n}  \sum_{k=1}^{K}
    \big| t_{i,k}-\mu \big|  f({t_{i,k}}|{\boldsymbol{\theta}}^{(s)}) \Delta t_i,
\label{EQ:LaplaceDirect}
\end{align}
where $\Delta t_i=t_{i,k}-t_{i,k-1}$, $t_{i,0}=a_i$ and $t_{i,K}=b_i$.
When this direct numerical integration is used,
the terms,  $f({t_{i,k}}|{\boldsymbol{\theta}}^{(s)})$ and $\Delta t_i$,
are involved inside the sum in equation~(\ref{EQ:LaplaceDirect}) and these are not constant. 
On the other hand, the QEM and MCEM algorithms do not include 
these $\Delta t_i=t_{i,k}-t_{i,k-1}$ terms.
Therefore, it can be easily seen that the median of $t_{i,k}$ can not be 
the maximizer of equation~(\ref{EQ:LaplaceDirect}) with respect to $\mu$. 
To the best of our knowledge, a closed-form maximizer for equation~(\ref{EQ:LaplaceDirect}) 
does not exist. As mentioned earlier, the use of the direct numerical integration makes it
very difficult or even impossible to find 
the closed-form maximizer in the M-step.
The point to be made here is that direct numerical integration is not useful 
because it is still requires an intractable or at the very least, extremely difficult,
maximization in the M-step. The advantage of MCEM and QEM over direct
numerical integration is that they simplify the M-step considerably.

\subsection{Rayleigh distribution}
Let the random variables $z_i$ be iid Rayleigh random variables 
with parameter $\beta$ whose pdf is given by 
\[
f(z|\beta)
  = \frac{z}{\beta^2} \exp\big(-\frac{z^2}{2\beta^2}\big),
\quad z>0,~ \beta>0.
\]
Using equation~(\ref{EQ:likelihood3}), we have the complete-data log-likelihood of 
$\beta$
\begin{align*}
\log L^c({{\beta}} | {\mathbf{z}})
&= C-2n\log\beta +\sum_{i=1}^{n}\log z_i - \frac{1}{2\beta^2}\sum_{i=1}^{n}z_i^2,
\end{align*}
where the pdf of the random variable $z_i$ is given by 
\begin{equation}  \label{EQ:RayleighTruncation}
 p_{z_i}(z|\beta) = \frac{ \frac{z}{\beta^2}\exp\big(-\frac{z^2}{2\beta^2}\big) }
                {\exp\big(-\frac{a_i^2}{2\beta^2}\big)-\exp\big(-\frac{b_i^2}{2\beta^2}\big)}
\end{equation}  
for $ a_i < z < b_i$. 
Similarly as before, if $a_i=b_i$, then the random variables $z_i$ 
degenerate at $z_i=a_i$. 

\begin{itemize}
\item \textsf{E-step:} \\
At the $s$-th step in the EM sequence denoted by $\beta^{(s)}$, we have
the expected log-likelihood
\begin{align*} 
&Q(\beta|\beta^{(s)})   \\
&\quad
  =\int\log L^c(\beta|{\mathbf{z}}) p({\mathbf{z}}|\beta^{(s)}) d{\mathbf{z}}\\
&\quad
  = C - 2n\log\beta  \\
&\qquad
   +\sum_{i=1}^{n} \int_{a_i}^{b_i}
    (\log z_i-\frac{1}{2\beta^2}z_i^2) p_{z_i}({z_i} | \beta^{(s)}) d{z_i}.
\end{align*}

The calculation of the above integration part does not have a closed form.
Using the MCEM, we have the approximate expected log-likelihood
\begin{align*}
\widehat{Q}({{\beta}}|{{\beta}}^{(s)}) 
&= \frac{1}{K}\sum_{k=1}^{K}
   \log L^c({{\beta}}|{\mathbf{z}}^{(k)})\\
&= C - 2n\log\beta +\frac{1}{K}\sum_{k=1}^{K}\sum_{i=1}^{n}\log z_{i,k}  \\
&\qquad -\frac{1}{2\beta^2}\frac{1}{K}\sum_{k=1}^{K} \sum_{i=1}^{n}z_{i,k}^2,
\end{align*}
where ${\mathbf{z}}^{(k)}=(z_{1,k},\ldots,z_{n,k})$ and $z_{i,k}$ for $k=1,2,\ldots,K$ are from
$p_{z_i}(z|{\beta}^{(s)})$ defined in equation~(\ref{EQ:RayleighTruncation}).

\bigskip
\item \textsf{M-step:} \\
We then obtain the following MCEM (or QEM) sequences by differentiating
$\widehat{Q}({{\beta}}|{{\beta}}^{(s)})$ 
\begin{equation}  \label{EQ:MCEMRayleigh}
\beta^{(s+1)} 
=\sqrt{\frac{1}{2n} \sum_{i=1}^n  \frac{1}{K} \sum_{k=1}^K z_{i,k}^2 }.
\end{equation}
\end{itemize}
In the above, if the quantiles $q_{i,k}$ are used instead of a random sample $z_{i,k}$, 
then the QEM sequences are obtained.

\subsection{Weibull distribution}
We assume that $X_i$ are iid Weibull random variables 
with the pdf and cdf given by  
$f(x) = \lambda \beta x^{\beta-1} \exp(-\lambda x^{\beta})$ and
$F(x) =  1 - \exp(-\lambda x^{\beta})$, respectively. 

Using equation~(\ref{EQ:likelihood3}),
we obtain the complete-data log-likelihood of 
$\boldsymbol{\theta}=(\lambda, \beta)$
\[
\log L^c( \boldsymbol{\theta} )
= \sum_{i=1}^{n}
  \Big\{ \log\lambda +  \log\beta + (\beta-1)\log z_i - \lambda z_i^{\beta} \Big\},
\]
where the pdf of $z_i$ is given by
\begin{equation}  \label{EQ:fz}
p_{z_i}(z | \boldsymbol{\theta} )
= \frac{ \lambda \beta z^{\beta-1} \exp(-\lambda z^{\beta}) }
       { \exp(-\lambda a_i^{\beta}) - \exp(-\lambda b_i^{\beta}) },
\end{equation}
for $a_i < z < b_i$.
Similarly as before, if $a_i=b_i$, then the random variables $z_i$ 
degenerate at $z_i=a_i$. 

\begin{itemize}
\item \textsf{E-step:} \\
Denote the estimate of $\boldsymbol{\theta}$ at the $s$-th EM sequence by
$\boldsymbol{\theta}^{(s)} = (\lambda^{(s)}, \beta^{(s)})$.
It follows from $Q( \boldsymbol{\theta} | \boldsymbol{\theta}^{(s)} ) 
= E\big[\log L^c(\boldsymbol{\theta}) \big]$ that
\begin{align*}
Q ( \boldsymbol{\theta} | \boldsymbol{\theta}^{(s)} ) 
&= n\log\lambda + n\log\beta + (\beta-1)\sum_{i=1}^{n} A_{i}^{(s)}  \\
& \qquad - \lambda \sum_{i=1}^{n} B_{i}^{
(s)},
\end{align*}
where
$A_{i}^{(s)} = E\big[ \log z_i | \boldsymbol{\theta}^{(s)} \big]$
and
$B_{i}^{(s)} = E\big[ z_i^\beta | \boldsymbol{\theta}^{(s)} \big]$.

\bigskip
\item \textsf{M-step:} \\
Differentiating
$Q(\lambda| \lambda^{(s)})$
with respect to $\lambda$ and $\beta$ and setting this to zero, we obtain
\begin{align}
\frac{\partial Q( \boldsymbol{\theta} | \boldsymbol{\theta}^{(s)} )}{\partial\lambda}
&= \frac{n}{\lambda} -  \sum_{i=1}^{n} B_{i}^{(s)}(\beta) = 0
\label{EQ:WeibullQlambda}  \\
\intertext{and}
\frac{\partial Q( \boldsymbol{\theta} | \boldsymbol{\theta}^{(s)} )}{\partial\beta}
&= \frac{n}{\beta} + \sum_{i=1}^{n} A_{i}^{(s)}
 - \lambda \sum_{i=1}^{n} \frac{\partial B_{i}^{(s)}(\beta)}{\partial\beta} = 0.
\label{EQ:WeibullQbeta} 
\end{align}
Solving equation~(\ref{EQ:WeibullQlambda}) for $\lambda$ and substituting this 
$\lambda$ into equation~(\ref{EQ:WeibullQbeta}),
we obtain the following expression involving $\beta$
\[
\frac{1}{\beta} + \frac{1}{n}\sum_{i=1}^{n} A_{i}^{(s)} -
 \frac{\sum_{i=1}^{n}\frac{\partial B_{i}^{(s)}(\beta)}{\partial\beta}}
 {\sum_{i=1}^{n} B_{i}^{(s)}(\beta)} = 0.
\]
Note that the $(s+1)$st element of EM sequence of $\beta$ is
the solution of the equation above. Therefore, after finding $\beta^{(s+1)}$, 
we can then obtain the $(s+1)$st element of the EM sequence of  $\lambda^{(s+1)}$
\[
\lambda^{(s+1)} = \frac{n}{\sum_{i=1}^{n} B_{i}^{(s)}(\beta^{(s+1)})}.
\]
\end{itemize}


Note that, in the Weibull case, it is extremely
difficult to obtain explicit expression for the expectations, 
$E\big[ \log z_i | \boldsymbol{\theta}^{(s)} \big]$
and $E\big[ z_i^\beta | \boldsymbol{\theta}^{(s)} \big]$
in the E-step. Fortunately, the quantile function of $z_i$ at the $s$-th 
step can be easily obtained, which makes the QEM particularly useful
in the case of the Weibull assumption. 
Specifically, based on equation~(\ref{EQ:fz}), we have
\begin{align*}
q_{i,k} 
&= F^{-1}_{z_i} (\xi_k | \boldsymbol{\theta}^{(s)} )  \\
&= \Bigg[
   -\frac{1}{\lambda^{(s)}} \log\Big\{ (1-\xi_k)\exp({-\lambda^{(s)} a_i^{\beta^{(s)}}}) \\
&\qquad\qquad + \xi_k \exp({-\lambda^{(s)} b_i^{\beta^{(s)}}})  \Big\}
\Bigg]^{1/\beta^{(s)}}.
\end{align*}
Using the above quantiles, we obtain the following QEM algorithm.
\begin{itemize}
\item \textsf{E-step:} \\
Denote the quantile approximation of $Q(\cdot)$ by $\widehat{Q}(\cdot)$.
Then, we have
\begin{align*}
&\widehat{Q}( \boldsymbol{\theta} | \boldsymbol{\theta}^{(s)} )  
= n\log\lambda + n\log\beta  \\
&\quad + (\beta-1)\sum_{i=1}^{n} \frac{1}{K} \sum_{k=1}^{K} \log q_{i,k} 
  - \lambda \sum_{i=1}^{n} \frac{1}{K} \sum_{k=1}^{K} q_{i,k}^\beta.
\end{align*}

\bigskip
\item \textsf{M-step:} \\
Differentiating
$\widehat{Q}(\lambda| \lambda^{(s)})$
with respect to $\lambda$ and $\beta$ and setting this to zero, we obtain
\begin{align}
\frac{\partial \widehat{Q}( \boldsymbol{\theta} | \boldsymbol{\theta}^{(s)} )}{\partial\lambda}
&= \frac{n}{\lambda}-\frac{1}{K} \sum_{i=1}^{n} \sum_{k=1}^{K} q_{i,k}^\beta =0
\label{EQ:WeibullQhatlambda} \\
\intertext{and}
\frac{\partial \widehat{Q}( \boldsymbol{\theta} | \boldsymbol{\theta}^{(s)} )}{\partial\beta}
&= \frac{n}{\beta} + \frac{1}{K} \sum_{i=1}^{n} \sum_{k=1}^{K} \log q_{i,k} \\
&- \lambda \frac{1}{K} \sum_{i=1}^{n}\sum_{k=1}^{K} q_{i,k}^\beta \log q_{i,k}=0.
\label{EQ:WeibullQhatbeta} 
\end{align}
Solving equation~(\ref{EQ:WeibullQhatlambda}) for $\lambda$ and substituting this 
$\lambda$ into equation~(\ref{EQ:WeibullQhatbeta}),
we have the equation of $\beta$
\begin{align}  \label{EQ:EEbeta}
\frac{1}{\beta}  
&+ \frac{1}{n K}\sum_{i=1}^{n} \sum_{k=1}^{K} \log q_{i,k} \notag  \\
&- \frac{\sum_{i=1}^{n} \sum_{k=1}^{K}  q_{i,k}^\beta \log q_{i,k}  }
 {\sum_{i=1}^{n} \sum_{k=1}^{K} q_{i,k}^\beta } = 0.
\end{align}
Note that the $(s+1)$st element of QEM sequence of $\beta$ is
the solution of the equation above. Therefore, after finding $\beta^{(s+1)}$,
we can then obtain the $(s+1)$st element of the QEM sequence of $\lambda^{(s+1)}$
\[
\lambda^{(s+1)} 
   = \frac{nK}{ \sum_{i=1}^{n} \sum_{k=1}^{K} q_{i,k}^{\beta^{(s+1)}} }.
\]
\end{itemize}

We should point out that, in the M-step,
we need to estimate the shape parameter $\beta$  by solving equation~(\ref{EQ:EEbeta}) 
numerically. 
Note that upper and lower bounds for the root of equation~(\ref{EQ:EEbeta}) can be
explicitly obtained. This implies that the solution can be obtained using
only a one dimensional root search and the uniqueness of the solution is
guaranteed.  Under mild conditions, we provide a proof of the uniqueness in the
appendix section along with the upper and lower bounds for $\beta$.

\section{Simulation study}
In order to examine the performance of the proposed QEM
method, we carry out two different simulations. In the first simulation,
we assume that the lifetimes are normally distributed. The second
simulation assumes that the lifetimes have a Rayleigh distribution.
The number of samples
used for the MCEM and QEM algorithms was varied so that 
$K=10$, $10^2$, $10^3$, and $10^4$.
The Monte Carlo simulations are based on $5,000$ replications.
The Monte Carlo simulations are performed using the R language.\cite{R:2018}

\begin{table*}[tb!]
\small\sf\centering
\caption{\label{MCEM:Simulation1}
Estimated biases and MSEs, and SREs of 
the EM, MCEM and QEM estimators assuming normally distributed data 
with $\mu=50$ and $\sigma=5$.}
\begin{tabular}{l@{\qquad}rrr}
\hline
       & \multicolumn{3}{c}{$\hat{\mu}$}  \\
Method & \multicolumn{1}{c}{Bias} & \multicolumn{1}{c}{MSE} & \multicolumn{1}{c}{SRE} \\
\hline\\[-1.5ex]
EM     & 1.342988$\times10^{-5}$ & 1.779955$\times10^{-10}$ &   ------                  \\[1ex]
MCEM \\
$K=10$   & 7.169276$\times10^{-2}$ & 8.381887$\times10^{-3}$  & 2.123573$\times10^{-8}$  \\
$K=10^2$ & 2.223300$\times10^{-2}$ & 8.170053$\times10^{-4}$  & 2.178633$\times10^{-7}$  \\
$K=10^3$ & 7.135135$\times10^{-3}$ & 8.417492$\times10^{-5}$  & 2.114590$\times10^{-6}$  \\
$K=10^4$ & 2.265630$\times10^{-3}$ & 8.433365$\times10^{-6}$  & 2.110610$\times10^{-5}$  \\[1ex]
QEM \\
$K=10$   & 2.511190$\times10^{-2}$ & 2.558357$\times10^{-5}$  & 6.957412$\times10^{-6}$  \\
$K=10^2$ & 2.382535$\times10^{-3}$ & 2.305853$\times10^{-7}$  & 7.719289$\times10^{-4}$  \\
$K=10^3$ & 2.349116$\times10^{-4}$ & 2.432084$\times10^{-9}$  & 7.318639$\times10^{-2}$  \\
$K=10^4$ & 3.232357$\times10^{-5}$ & 2.176558$\times10^{-10}$ & 8.177841$\times10^{-1}$  \\
\hline\hline
      & \multicolumn{3}{c}{$\hat{\sigma}$} \\
Method& \multicolumn{1}{c}{Bias} & \multicolumn{1}{c}{MSE} & \multicolumn{1}{c}{SRE} \\
\hline\\[-1.5ex]
EM       & 3.706033$\times10^{-5}$ & 1.143094$\times10^{-10}$ &   ------              \\[1ex]
MCEM \\
$K=10$   & 1.139404$\times10^{-1}$ & 2.133204$\times10^{-2}$  & 5.358577$\times10^{-9}$ \\
$K=10^2$ & 3.540433$\times10^{-2}$ & 2.069714$\times10^{-3}$  & 5.522955$\times10^{-8}$ \\
$K=10^3$ & 1.137090$\times10^{-2}$ & 2.130881$\times10^{-4}$  & 5.364419$\times10^{-7}$ \\
$K=10^4$ & 3.621140$\times10^{-3}$ & 2.150602$\times10^{-5}$  & 5.315227$\times10^{-6}$ \\[1ex]
QEM \\
$K=10$   & 5.580507$\times10^{-2}$ & 1.272262$\times10^{-4}$  & 8.984739$\times10^{-7}$ \\
$K=10^2$ & 5.890585$\times10^{-3}$ & 1.418158$\times10^{-6}$  & 8.060413$\times10^{-5}$ \\
$K=10^3$ & 6.315578$\times10^{-4}$ & 1.644996$\times10^{-8}$  & 6.948917$\times10^{-3}$ \\
$K=10^4$ & 9.699447$\times10^{-5}$ & 4.529568$\times10^{-10}$ & 2.523627$\times10^{-1}$ \\
\hline
\end{tabular}
\end{table*}

We illustrate the performance of the proposed method with
the EM and MCEM estimators by computing the respective mean biases and
mean square errors (MSEs). The bias is defined as the sample average of
the differences between the estimates under consideration and the MLE.
The MLE is obtained by solving the log-likelihood estimating equation numerically
using the \texttt{nlm()} function in R.
The MSE is defined as the sample average of the squares of
the differences between the estimates under consideration and the MLE.

Note that in order to compare the efficiency of the MCEM algorithm and 
QEM algorithms, we used an equal and fixed number of iterations 
in both simulations. In this manner, we compare the
accuracy when the computational burden of each algorithm is the same.
Both algorithms were stopped after ten iterations ($s=10$) and the simulation
results are shown in Tables~\ref{MCEM:Simulation1} and \ref{MCEM:Simulation2}.
Rather than fixing the number of iterations, we could have
taken the alternative route of using the same stopping criteria for both
the QEM and MCEM algorithms. 
Clearly, if the QEM accuracy is greater with the iterations being fixed, 
then a stopping criteria methodology would
lead to similar accuracy but a greater number of iterations would be
required for the MCEM stopping criteria to be triggered. So, the two
comparison methodologies  are, for all intents and purposes, equivalent
and we chose the methodology in which the number of iterations are
fixed to the same pre-determined value for both the MCEM and the QEM.

In the first simulation, a random sample of size $n=20$ was 
generated from the normal distribution with $\mu=50$ and $\sigma=5$.
Also, the largest five data points from the sample were assumed to be right-censored. 
In order to  compare the MCEM and QEM algorithms 
with the EM algorithm as a reference,
a univariate statistical dispersion measure based on the MSE 
can be used to compare algorithm efficiency. 
Analogous to the relative 
efficiency,\cite{Lehmann:1999,Park/Leeds:2016,Park/Ouyang/Byun/Leeds:2017}
the simulated relative efficiency (SRE) is defined as 
\[
\mathrm{SRE} = \frac{\mathrm{simulated~MSE~of~the~EM~estimator}}
                    {\mathrm{simulated~MSE~under~consideration}}.
\]

\begin{table}
\small\sf\centering
\caption{\label{MCEM:Simulation2}
Estimated biases and MSEs of the MCEM and QEM estimators
assuming Rayleigh distributed data with $\beta=10$.}
\begin{tabular}{lcrr}
\hline  
&&\multicolumn{2}{c}{$\hat{\beta}$}  \\
Method && \multicolumn{1}{c}{Bias} & \multicolumn{1}{c}{MSE} \\
\hline 
MCEM  \\
$K=10$   &&  0.1457421790  &   3.419463$\times\!10^{-2}$   \\
$K=10^2$ &&  0.0450846748  &   3.260372$\times\!10^{-3}$   \\
$K=10^3$ &&  0.0142957976  &   3.283920$\times\!10^{-4}$   \\
$K=10^4$ &&  0.0045336330  &   3.269005$\times\!10^{-5}$   \\
QEM  \\                                             
$K=10$   &&  0.0560471712  &   5.554717$\times\!10^{-5}$   \\
$K=10^2$ &&  0.0057055322  &   5.739379$\times\!10^{-7}$   \\
$K=10^3$ &&  0.0005675668  &   5.517117$\times\!10^{-9}$   \\
$K=10^4$ &&  0.0000527132  &   3.759086$\times\!10^{-11}$  \\
\hline 
\end{tabular} 
\end{table}

By comparing the efficiencies in Table~\ref{MCEM:Simulation1},
it is clear that the EM algorithm is as efficient as the MLE. More importantly, the
Table~\ref{MCEM:Simulation1} indicates that the QEM results 
in smaller MSE and much greater efficiency compared to that of the MCEM. 
For example, using $K=10,000$, the SRE of the MCEM is only 
$2.110610\times10^{-5}$ for $\hat{\mu}$ 
and $5.315227\times10^{-6}$ for $\hat{\sigma}$
while the SRE of the QEM is  $8.177841\times10^{-1}$ for $\hat{\mu}$ and 
$2.523627\times10^{-1}$ 
for $\hat{\sigma}$.
Strikingly, the QEM using only $K=100$ clearly outperforms
the MCEM using $K=10,000$.

In the second simulation, we draw a random sample
of size $n=20$ from the Rayleigh distribution with $\beta=10$. Just as
was the case in the first simulation, we assume that the five largest
data points from the sample were right-censored. The results are shown
in Table~\ref{MCEM:Simulation2}. 
Note that in this case we can only compare the MCEM and the
QEM because the EM algorithm cannot be implemented due to its extremely
complex E-step. Therefore, the SREs are excluded from Table~\ref{MCEM:Simulation2}.
As expected based on the E-step accuracy results developed earlier, 
the results in Table~\ref{MCEM:Simulation2} illustrate that the QEM outperforms the MCEM.
For example, the MSE of the QEM with only $K=10$ is quite
comparable to that of the MCEM with a random sample of size $K=10,000$.
This is understandable given that E-step
accuracy of the QEM in this particular case is $O(1/K^2)=O(1/100)$ with $K=10$ 
and the E-step accuracy of the MCEM is $O_p(1/\sqrt{K})=O_p(1/100)$ with $K=10,000$. 

Another way of comparing the accuracies of the QEM and MCEM is to 
consider the ratio of the respective MSEs for a given value of $K$.
Using the results in Tables~\ref{MCEM:Simulation1} and \ref{MCEM:Simulation2},
we calculated the following ratio for each of $K=10, 10^2, 10^3, 10^4$ 
in Table~\ref{TBL:MSEcomparison}:
\[
\frac{\textrm{MSE (MCEM)}}{\textrm{MSE (QEM)}}. 
\]
Table~\ref{TBL:MSEcomparison} clearly shows that the MSE of the QEM is much smaller than 
that of the MCEM for a given value of $K$.

\begin{table}
\small\sf\centering
\caption{
\label{TBL:MSEcomparison}
Ratios of MSEs, $\textrm{MSE(MCEM)}/\textrm{MSE(QEM)}$.}
\begin{tabular}{lcrrr}
\hline 
$K$ && $\hat{\mu}$ & $\hat{\sigma}$  & $\hat{\beta}$ \\
\hline 
$K=10$   &&    327.6 &   167.7 &    615.6  \\
$K=10^2$ &&   3543.2 &  1459.4 &   5680.7  \\
$K=10^3$ &&  34610.2 & 12953.7 &  59522.4  \\
$K=10^4$ &&  38746.3 & 47479.2 & 869627.6  \\
\hline 
\end{tabular} 
\end{table}

Next, the identical simulations for the normal
and Rayleigh cases were carried out again in order to compare both the CPU
and real time performance of the QEM and MCEM algorithms. Since, in the
normal distribution case, the accuracy of the QEM using $K=100$ is already known
to be quite comparable to that of the MCEM using $K=1,000$, 
these respective values of $K$ were used again. 
In the Rayleigh distribution case,  the accuracy of the QEM with $K=10$ is quite comparable to 
that of the MCEM with $K=10,000$, so we used these  respective
values of $K$ were used. 
The running time of the algorithms is easily
measured through the use of the \texttt{proc.time()} function in R.
This  \texttt{proc.time()} function reports user, system and elapsed times. 
The \emph{user time} is the CPU time charged for the execution of  the calling process, 
the \emph{system time} is the CPU time charged for execution by the system 
on behalf of the calling process, and  
the elapsed time is the \emph{real} elapsed time since the process was started.
For more details regarding the \texttt{proc.time()} function,
one is referred to its help page in R.
The simulations for the running times were carried out 
using a Ubuntu Linux workstation with
Intel Core i7-7700K CPU. The results are summarized in Table~\ref{TBL:CPUcomparison} and
they indicate that the computations used in QEM algorithm take much
less time than those used in the MCEM algorithm.

\begin{table}
\small\sf\centering
\caption{\label{TBL:CPUcomparison}Comparison of running times (in seconds).}
\begin{tabular}{lcrrr}
\hline 
Method && User &  System  & Elapsed \\
\hline   
&& \multicolumn{3}{c}{Normal distribution} \\
MCEM  && 529.545  & 3.367  & 532.922  \\
QEM   &&   5.058  & 0.000  &   5.059  \\
\hline
&& \multicolumn{3}{c}{Rayleigh distribution} \\
MCEM  && 200.949 & 3.204  & 204.159  \\
QEM   &&   1.982 & 0.000  &   1.982  \\
\hline 
\end{tabular}
\end{table}

\section{Examples of application of the proposed methods}
In this section, we provide four numerical examples of parameter
estimation using data sets from the literature in addition 
to artificially generated data sets.
The parameters are estimated using the EM (when available),
MCEM and QEM algorithms.

\begin{table}
\small\sf\centering
\caption{\label{TBL:MCEM:norm1}Iterations of the EM, MCEM, and QEM sequences 
using data from Gupta.\cite{Gupta:1952}}
\begin{tabular}{rrrr}
\hline 
Step  &\multicolumn{3}{c}{$\mu^{(s)}$} \\
$s$   &\multicolumn{1}{c}{EM}&\multicolumn{1}{c}{MCEM}&\multicolumn{1}{c}{QEM} \\
\hline   
0 & 0      &      0 &      0 \\
1 & 1.8467 & 1.8456 & 1.8467  \\
2 & 1.8058 & 1.8074 & 1.8057  \\
3 & 1.7761 & 1.7771 & 1.7760  \\
4 & 1.7593 & 1.7597 & 1.7593  \\
5 & 1.7504 & 1.7503 & 1.7503  \\
6 & 1.7459 & 1.7458 & 1.7459  \\
7 & 1.7439 & 1.7440 & 1.7439  \\
8 & 1.7429 & 1.7428 & 1.7429  \\
9 & 1.7425 & 1.7422 & 1.7425  \\
10& 1.7424 & 1.7421 & 1.7424  \\
\hline 
Step &\multicolumn{3}{c}{$\sigma^{(s)}$} \\
$s$  &\multicolumn{1}{c}{EM}&\multicolumn{1}{c}{MCEM}&\multicolumn{1}{c}{QEM} \\
\hline 
0 &      1   &      1   &      1 \\
1 & 0.2968   &  0.2973  &  0.2966 \\
2 & 0.1931   &  0.1959  &  0.1930 \\
3 & 0.1370   &  0.1386  &  0.1369 \\
4 & 0.1070   &  0.1076  &  0.1069 \\
5 & 0.0919   &  0.0919  &  0.0919 \\
6 & 0.0848   &  0.0847  &  0.0848 \\
7 & 0.0816   &  0.0816  &  0.0816 \\
8 & 0.0802   &  0.0802  &  0.0802 \\
9 & 0.0796   &  0.0792  &  0.0796 \\
10& 0.0793   &  0.0789  &  0.0793 \\
\hline 
\end{tabular} 
\end{table}

\subsection{Censored normal data}
First, consider the data presented earlier by Gupta\cite{Gupta:1952} 
in which the largest three out of the $n=10$ observations have been censored.
The Type-II right-censored observations are therefore: 
1.613, 1.644, 1.663, 1.732, 1.740, 1.763, 1.778, $1.778^+$, $1.778^+$, $1.778^+$.

The MLEs of $\mu$ and $\sigma$ are $\hat{\mu}=1.742$ and
$\hat{\sigma}=0.079$.  We also generate the EM sequences from
equations~(\ref{EQ:EMnormal1}) and (\ref{EQ:EMnormal2}) in order to compare these
estimates with the MLE. The starting values used for the EM algorithm
were $\mu^{(0)}=0$ and ${\sigma^2}^{(0)}=1$.  Similarly, we generate the
MCEM sequences from equations~(\ref{EQ:MCEMnormal1}) and (\ref{EQ:MCEMnormal2}) in
order to obtain the MCEM and QEM estimates. The MCEM and QEM algorithms
were run using $K=1,000$ and the algorithms were stopped after ten 
iterations.  Table~\ref{TBL:MCEM:norm1} illustrates the results for
all three algorithms.  Note that the EM algorithm estimate is identical
to the MLE up to the third decimal point after nine iterations.  
Also, as would be expected on the theoretical
convergence properties developed earlier, 
the QEM estimate is much closer to the MLE and the EM
estimate than the MCEM estimate.

\subsection{Censored Laplace data}
Next, we consider the data presented earlier 
by Balakrishnan\cite{Balakrishnan:1996} in which,
out of $n=20$ observations, the largest two have been censored.
The Type-II right-censored observations thus obtained are:  
 32.00692, 37.75687, 43.84736, 46.26761, 46.90651, 
 47.26220, 47.28952, 47.59391, 48.06508, 49.25429,
 50.27790, 50.48675, 50.66167, 53.33585,
 53.49258, 53.56681, 53.98112, 54.94154, $54.94154^+$, $54.94154^+$.

\begin{table}[t!]
\small\sf\centering
\caption{\label{MCEM:Laplace} Iterations of the MCEM and QEM sequences
using data from Balakrishnan.\cite{Balakrishnan:1996}}
\begin{tabular}{rrrcrr}
\hline 
& \multicolumn{2}{c}{$\mu^{(s)}$}
&&\multicolumn{2}{c}{$\sigma^{(s)}$} \\
\cline{2-3}\cline{5-6} \\[-1ex]
$s$ & \multicolumn{1}{c}{MCEM} & \multicolumn{1}{c}{QEM}
&&    \multicolumn{1}{c}{MCEM} & \multicolumn{1}{c}{QEM} \\
\hline   
0 &         0 &        0 &&       1  &       1 \\
1 &  49.76609 & 49.76609 && 4.320983 & 4.318817 \\
2 &  49.76609 & 49.76609 && 4.669010 & 4.650584 \\
3 &  49.76609 & 49.76609 && 4.669581 & 4.683749 \\
4 &  49.76609 & 49.76609 && 4.682357 & 4.687064 \\
5 &  49.76609 & 49.76609 && 4.693247 & 4.687395 \\
6 &  49.76609 & 49.76609 && 4.687793 & 4.687429 \\
7 &  49.76609 & 49.76609 && 4.693793 & 4.687432 \\
8 &  49.76609 & 49.76609 && 4.678954 & 4.687432 \\
9 &  49.76609 & 49.76609 && 4.702827 & 4.687432 \\
10&  49.76609 & 49.76609 && 4.671909 & 4.687432 \\
\hline 
\end{tabular} 
\end{table}

In this case, Balakrishnan\cite{Balakrishnan:1996}
computed the best linear unbiased estimates (BLUE) of $\mu$ and $\sigma$
and  obtained  $\hat{\mu}=49.56095$ and $\hat{\sigma}=4.81270$.
The MLE is $\hat{\mu}=49.76609$ and $\hat{\sigma}=4.68761$.

We also generated the MCEM sequences from equations~(\ref{EQ:MCEMLaplace1}) and
(\ref{EQ:MCEMLaplace2}) in order to compute the MCEM and QEM estimates.
Both algorithms were run with $K=1,000$ for ten iterations with starting
values $\mu^{(0)}=0$ and $\sigma^{(0)}=1$.
The iterations associated with the MCEM and QEM algorithms are
shown in Table~\ref{MCEM:Laplace}.
As was expected, the QEM estimate is significantly
closer to the MLE than the MCEM estimate, particularly with respect
to $\sigma$. We should also note that both the MCEM and QEM estimates are
closer to the MLE than the BLUE.

\subsection{Censored Rayleigh data}
Next, we generated a random sample of $n=20$ from the Rayleigh distribution with 
$\beta=5$ and the five largest data points were considered to be right-censored.
The Type-II right-censored observations thus obtained are:
1.950, 2.295, 4.282, 4.339,  4.411, 4.460, 4.699, 5.319, 5.440, 5.777,
7.485, 7.620, 8.181, 8.443, 10.627, $10.627^+$, $10.627^+$, $10.627^+$, $10.627^+$, $10.627^+$.

\begin{table}[tb!]
\small\sf\centering
\caption{\label{MCEM:Rayleigh}
Iterations of the MCEM and QEM sequences using simulated data set
from the Rayleigh distribution.}
\begin{tabular}{crrcrr}
\hline 
    & \multicolumn{2}{c}{${\beta}^{(s)}$}
   && \multicolumn{2}{c}{${\beta}^{(s)}$} \\
\cline{2-3} \cline{5-6} \\[-1ex]
$s$ & \multicolumn{1}{c}{MCEM} & \multicolumn{1}{c}{QEM}
   && \multicolumn{1}{c}{MCEM} & \multicolumn{1}{c}{QEM} \\
\hline   
0 &    1      &  1      &&   10     & 10    \\
1 &    5.3363 &  5.3358 &&   7.3335 &  7.2946 \\
2 &    5.9395 &  5.9444 &&   6.4458 &  6.4435 \\
3 &    6.0888 &  6.0870 &&   6.2167 &  6.2126 \\
4 &    6.1170 &  6.1221 &&   6.1488 &  6.1536 \\
5 &    6.1413 &  6.1309 &&   6.1494 &  6.1387 \\
6 &    6.1336 &  6.1330 &&   6.1356 &  6.1350 \\
7 &    6.1214 &  6.1336 &&   6.1219 &  6.1341 \\
8 &    6.1290 &  6.1337 &&   6.1291 &  6.1338 \\
9 &    6.1261 &  6.1338 &&   6.1261 &  6.1338 \\
10&    6.1292 &  6.1338 &&   6.1292 &  6.1338 \\
\hline 
\end{tabular}
\end{table}

We then generated the MCEM and QEM sequences from equation~(\ref{EQ:MCEMRayleigh})
in order to compute the MCEM and QEM estimates.  
Both algorithms were run with $K = 1,000$ for ten 
iterations with two different starting values, namely $\beta^{(0)}=1$ and
$\beta^{(0)}=10$.
The iterations of the MCEM and QEM sequences are shown 
in Table~\ref{MCEM:Rayleigh}. 
The iteration sequences illustrate the difference in the rate of
convergence of the MCEM and QEM algorithms with the latter converging
extremely quickly. Note that the MLE is $\hat{\beta}=6.1341$ and the QEM
sequences are identical to the MLE up to the third decimal place after
the sixth iteration.

\subsection{Weibull interval-censored data}
The previous examples illustrated that the QEM algorithm
outperforms the MCEM both in terms of accuracy and rate of convergence.
In this example, we consider a real-data example of 
intermittent inspection of cracked parts. 
This part-cracking data set in this example was originally provided 
by Nelson\cite{Nelson:1982}
and has since then been widely used for illustration in the engineering 
literature and software.\cite{Kim/Yum:2000,SASQC:2013,ReliaSoft:2015}
The 167 identical parts in a machine
were intermittently inspected to obtain the number of
cracked parts in each interval. 
 The data from intermittent 
inspection are referred to as grouped data where only the number of failures
in each inspection are provided. The data represent cracked parts and
is provided in Table~\ref{TBL:Cracking}.  Other examples of grouped and
censored data can also be found in the statistics and engineering 
literature.\cite{Seo/Yum:1993,Shapiro/Gulati:1998,Xiong/Ji:2004,Meeker:1986,%
Nelson:1990,Sun:2006,Lee/Park:2006a,Park/Park/Cho/Bahn:2017}
These censored and grouped data can also 
be regarded as interval-censored data. 
Thus, the proposed method can be easily applicable to these data. 
Note that Seo and Yum\cite{Seo/Yum:1993} and Shapiro and Gulati\cite{Shapiro/Gulati:1998} 
have given an approximation of the MLE under the exponential distribution only.  
\begin{table}
\small\sf\centering
\caption{\label{TBL:Cracking}Observed frequencies of intermittent inspection data.} 
\begin{tabular}{r@{~$\sim$~}rlc}
\hline 
\multicolumn{2}{c}{Inspection} && Observed  \\
\multicolumn{2}{c}{time}       && failures  \\
\cline{1-2} \cline{4-4} 
    0 &  6.12 &&  5    \\
 6.12 & 19.92 && 16    \\
19.92 & 29.64 && 12    \\
29.64 & 35.40 && 18    \\
35.40 & 39.72 && 18    \\
39.72 & 45.24 &&  2    \\
45.24 & 52.32 &&  6    \\
52.32 & 63.48 && 17    \\
63.48 &       && 73    \\
\hline 
\end{tabular}
\end{table}

From Table~\ref{TBL:Cracking}, it becomes obvious
that these grouped data can be viewed as interval-censored data so
that the proposed QEM algorithm can be used to estimate the distribution
parameters. The QEM algorithm was used on this data set.
First, assuming that the data were exponentially distributed,
the QEM algorithm was applied.  Then, the QEM algorithm was run again assuming that
the data had a Weibull distribution.
In both cases, a stopping criterion was used
with $\epsilon=10^{-5}$ and the starting values used 
were $\lambda_0=1$ (exponential) and $\lambda_0=1$ and $\beta_0=1$ (Weibull).
In the first case, the exponential rate parameter $\lambda$ was  estimated as
$\hat{\lambda}=0.01209699$.
In the second case, the Weibull parameters were estimated as 
$\hat{\lambda}=0.001674018$ and $\hat{\beta}=1.497657$.

\section{Concluding remarks}
In this paper, we have illustrated that the QEM algorithm offers clear
advantages over the MCEM algorithm. 
The E-step accuracy of the QEM was shown to be $O(1/K^2)$ 
while that of  the MCEM was shown to be $O_p(1/\sqrt{K})$.
Thus, compared to the MCEM, the QEM reduces the computational complexity 
significantly for a given value of $K$. 
Also, the QEM possesses more stable convergence properties because the E-step
of the QEM has the accuracy of deterministic order while 
that of the MCEM has the accuracy of probabilistic order. 
The QEM algorithm provides 
a flexible and useful alternative for problems
where the E-step of the EM algorithm is either extremely complex or
completely intractable. Several examples were provided which illustrate
the usefulness of the proposed QEM algorithm.

\section*{{Acknowledgements}\label{SEC:ACK}}
A part of the simulations was performed on the Fedora Linux workstation system 
in Department of Mathematical Sciences at Clemson University 
while the author has worked at Clemson University.

This paper is dedicated to the memory and honor of Professor Byung Ho
Lee of Nuclear Engineering at Seoul National University.
He is a man of warmth and a major contributor to the development of
acoustics, creep and fatigue theory as well as nuclear engineering.
The author's interests in mathematics and engineering were formed under
his strong influence.  Professor Lee passed away in July, 2001.

This work was  supported in part by the National Research Foundation
of Korea (NRF) grant funded by the Korea government (NRF-2017R1A2B4004169).

\bibliographystyle{unsrt}


\appendix
\section*{Appendix}
\subsection*{Sketch of proof of the uniqueness and bounds of the Weibull shape parameter}
Following the approach used in Farnum and Booth\cite{Farnum/Booth:1997} 
and Park and Padgett,\cite{Park/Padgett:2006c}
the uniqueness of the solution in equation~(\ref{EQ:EEbeta}) can be proven as follows.
For convenience, we let 
\begin{align*}
 g(\beta)
 &= \frac{1}{\beta} 
\intertext{and} 
 h(\beta)
 &= \frac{ \sum_{i=1}^{n} \sum_{k=1}^{K} q_{i,k}^{\beta}\log q_{i,k}}{\sum_{i=1}^{n}\sum_{k=1}^{K} q_{i,k}^{\beta}}
  - \frac{1}{nK} \sum_{i=1}^{n}\sum_{k=1}^{K}  \log q_{i,k}  .
\end{align*}
Then solving equation~(\ref{EQ:EEbeta}) is equivalent to solving $g(\beta) = h(\beta)$.
We have 
\[
\frac{\partial h(\beta)}{\partial\beta}
=  \frac{A\cdot B - C^2}{ \{\sum_{i=1}^{n}\sum_{k=1}^{K} q_{i,k}^{\beta}\}^2 } , 
\]
where
\begin{align*}
A &= \sum_{i=1}^{n}\sum_{k=1}^{K} q_{i,k}^{\beta}\log^2 q_{i,k} \\
B &= \sum_{i=1}^{n}\sum_{k=1}^{K} q_{i,k}^{\beta} \\
\intertext{and} 
C &= \sum_{i=1}^{n}\sum_{k=1}^{K} q_{i,k}^{\beta}\log q_{i,k} .
\end{align*}
It is immediate from the  Jensen's inequality that 
\[
A\cdot B - C^2 \ge 0.
\]
Thus, we have ${\partial h(\beta)}/{\partial\beta} \ge 0$ so $h(\beta)$ 
is always increasing.
Since  $g(\beta)$ is strictly decreasing from $\infty$ to $0$ on $\beta \in [0,\infty]$, 
it suffices to show that $h(\beta)>0$ for some $\beta$.
Since
\[
\lim_{\beta\to\infty} h(\beta)
 = \frac{1}{nK} {\sum_{i=1}^{n}}\sum_{k=1}^{K} \Big\{\log q_{\max}-\log q_{i,k}\Big\},
\]
where $q_{\max} = \max_{i,k}\big\{ q_{i,k} \big\}$,
we have $h(\beta)>0$ for some $\beta$ 
unless $q_{i,k} = q_{\max}$ for all $i$ and $k$.
This condition is extremely unrealistic in practice.

Next, we provide upper and lower bounds of $\beta$.
These bounds guarantee that there is a unique
solution in the interval. Therefore, any root search type algorithm
will arrive at the solution in a stable manner. First we develop the
lower bound $\beta_L$.
 Clearly, since $h(\beta)$ is increasing, we have 
$ g(\beta) \le \lim_{\beta\to\infty} h(\beta)$ so that  
\[
\beta \ge \frac{ n K }{\sum_{i=1}^{n} \sum_{k=1}^{K} (\log q_{\max} - \log q_{i,k})}.
\]
The upper bound follows from the lower bound result.
Since $h(\beta)$ is again increasing, we have
$g(\beta)=h(\beta) \ge h(\beta_L)$, which leads to
\[
\beta \le \frac{ 1 }{h(\beta_L)}.
\]

\end{document}